\documentclass[
  reprint,  
  amsmath,amssymb,aps,showpacs,longbibliography,superscriptaddress
nofootinbib, showkeys
]{revtex4-1}
\usepackage{amssymb}
\usepackage{graphicx}
\usepackage{booktabs}
\usepackage{dcolumn}
\usepackage{bm}
\usepackage{mathrsfs}
\usepackage{mathtools}
\usepackage{amsmath}
\usepackage{amssymb}
\usepackage{epstopdf}
\usepackage{comment}
\usepackage{units}
\usepackage{mathtools}
\usepackage{dsfont}
\usepackage{upgreek}
\usepackage[font=footnotesize]{caption}
\usepackage{subfig}
\begin{document}
\preprint{APS/123-QED}

\title{ Stochastic line integrals and stream functions as metrics of irreversibility and heat transfer}

\author{Stephen Teitsworth}
\affiliation{Duke University, Department of Physics, Box 90305
Durham, NC 27708-0305}

\author{John C. Neu}
\affiliation{University of California at Berkeley, Department of Mathematics}

\date{\today}

\begin{abstract}

Stochastic line integrals provide a useful tool for quantitatively characterizing irreversibility and detailed balance violation in noise-driven dynamical systems. A particular realization is the stochastic area, recently studied in coupled electrical circuits. Here, we provide a general framework for understanding properties of stochastic line integrals and clarify their implementation for experiments and simulations. For two-dimensional systems, stochastic line integrals can be expressed in terms of a stream function, the sign of which determines the orientation of steady-state probability currents. Additionally, the stream function permits analytical understanding of the scaling dependence of stochastic area on key parameters such as the noise strength for both nonlinear and linear springs. Theoretical results are supported by numerical studies of an overdamped, two-dimensional mass-spring system driven out of equilibrium.
\end{abstract}

\maketitle

\section{Introduction}
\label{sec:intro}

Understanding and quantitatively characterizing irreversibility and related phenomena is a central task in the study of noise-driven non-equilibrium systems.  Such systems are of interest throughout the natural sciences with examples in diverse fields such as biophysics \cite{Battle_Science_2016, Gnesotto_2018}, climate dynamics \cite{Weiss_PRE_2007, Weiss_JSP_2019}, optically levitated nanoparticles \cite{Gieseler_2014, Millen_2014,GonzalezBallestero_Science_2021}, and electronic transport systems and circuits \cite{Bomze_PRL_2012, Gonzalez_PRE_2019, Teitsworth_2019}.  Well-known approaches for understanding non-equilibrium behavior focus on characterizing the dynamics with physically inspired metrics such as entropy production and heat transfer \cite{Ciliberto_PRL_2013, Ciliberto_JSM_2013b, Gnesotto_2018}.  One advantage of such techniques is that they connect directly to the non-equilibrium behavior of the underlying system, e.g., heat transfer from a hot mass to cold mass in a coupled mass-spring system \cite{Fogedby_2012, Ciliberto_JSM_2013b}.  However, for many systems of interest, the non-equilibrium dynamics is not directly determined by thermal gradients \cite{Battle_Science_2016, Gingrich_2019}, and the implementation of such metrics may not be straightforward.  For example, biophysical systems such as beating flagella and cilia are typically driven by chemical (metabolic and enzymatic) processes \cite{Battle_Science_2016}, while nonlinear electronic circuits may be driven by external voltage noise sources as well as internal non-thermal sources (e.g., shot noise) associated with nonlinear elements such as tunnel diodes \cite{Teitsworth_2019}.  For such systems, measured time series oftentimes do not have a direct or obvious connection to heat transport or entropy production, and the question naturally arises how to best quantify non-equilibrium dynamical behavior.  

This has motivated the introduction of a variety of theoretical and empirical metrics to characterize and quantify the extent of non-equilibrium behavior.  Such metrics include probability angular momentum \cite{Zia_Schmittmann_2007, Mellor_EPL_2016, Weiss_JSP_2019}, cycling frequencies \cite{Gradziuk_2019}, dissipation rate inference schemes \cite{Gingrich_2019, Gnesotto_2020}, information theoretic methods \cite{Frishman_2020}, and stochastic area \cite{Ghanta2017}.  One advantage of these metrics is that they can often be computed directly from experimental data time series and without knowledge of a detailed underlying model. In this paper, we introduce and explore two such metrics: 1) the stochastic line integral (SLI) which is a generalization of the recently introduced stochastic area \cite{Ghanta2017}, and 2) a stochastic stream function applicable to two-dimensional systems in steady state.

\begin{figure}
\centerline{\includegraphics[width=0.40\textwidth]{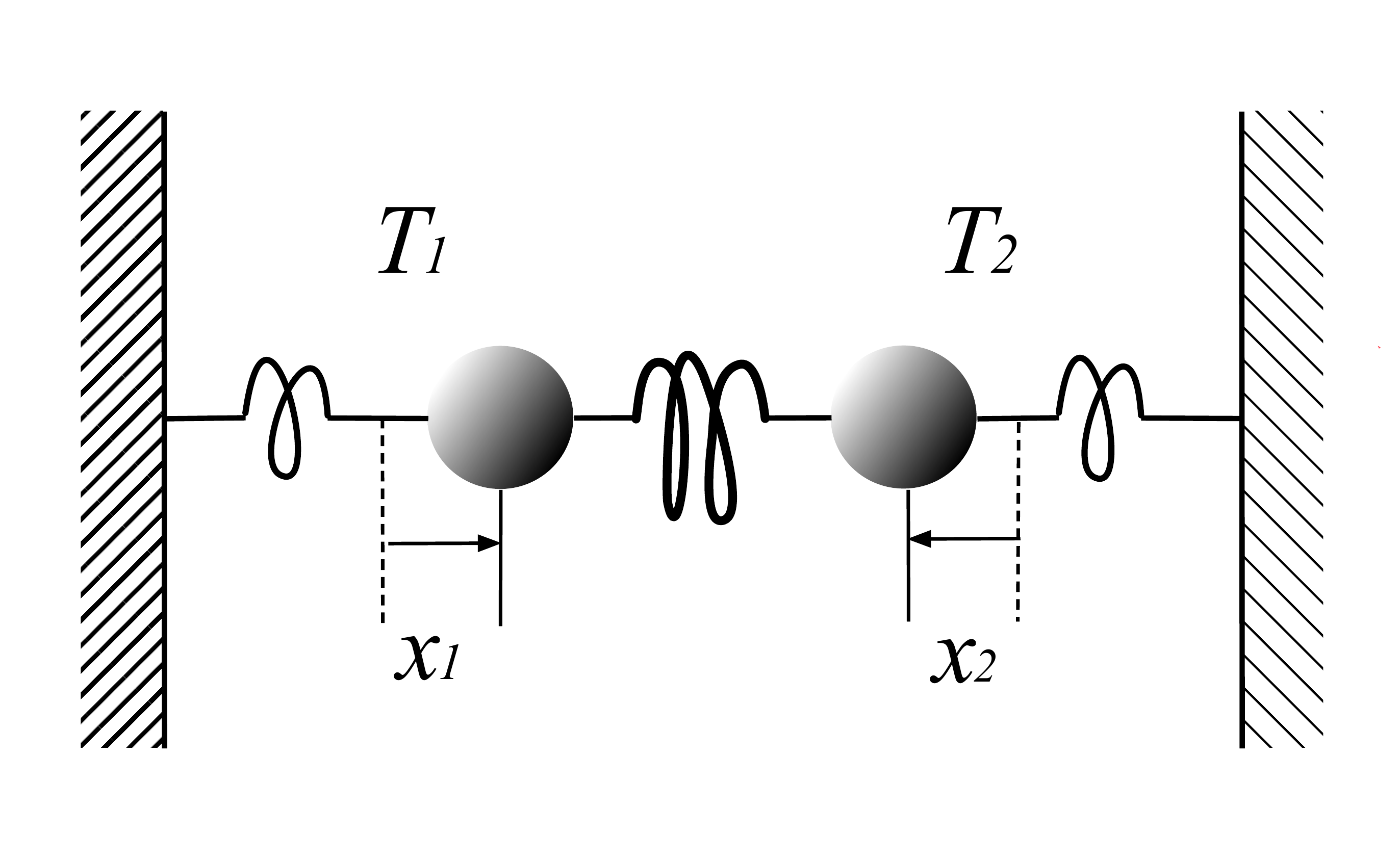}}
\caption{Linear mass-spring network of two Brownian particles (labelled 1 and 2) at respective temperatures $T_1$ and $T_2$.  ST needs to update this!}
\label{fig1}
\end{figure}

To frame the discussion, it is useful to start by recalling a key aspect of Brownian motion as a prototype for a large class of stochastic dynamical systems:  a small number of \textit{macroscopic} degrees of freedom
are distinguished from a great multitude of others called \textit{microscopic} by the nature of couplings.  The couplings of any one microscopic degree of freedom to the macroscopic degrees are small, but the collective action of the microscopic degrees upon the macroscopic degrees is comparable to the couplings of the macroscopic degrees among themselves.  This being so, individual identities of the microscopic degrees of freedom are not retained.  Their collective actions upon the macroscopic degrees of freedom are modeled by random \textit{fluctuations} and by \textit{dissipation}. As a concrete example, we can imagine a mechanical assembly immersed in a fluid, its movements stimulated by individual impacts of fluid particles, and inhibited by drag.  An example with two masses coupled by springs - which we analyze in greater detailed below - is depicted in Fig. \ref{fig1}.  

In the \textit{overdamped} limit which neglects inertia, we model the statistics of the macroscopic degrees of freedom by an \^{I}to stochastic differential equation (SDE) \cite{Gardiner_2009},
\begin{equation}
d\mathbf{x}(t) = \mathbf{u}(\mathbf{x}(t)) dt +  \sigma d\mathbf{W}(t) + O((dt)^{\frac{3}{2}}).
\label{eq:SDE_general}
\end{equation}
Here, $\mathbf{x}(t)$ is the time series of the state vector in $\mathbb{R}^n$, $dt$ is the timestep, and $d\mathbf{x}(t)$ is the \textit{forward difference}
\begin{equation}
d\mathbf{x}(t) := \mathbf{x}(t + dt) - \mathbf{x}(t).
\label{eq:forward_diff}
\end{equation}
The vector field $\mathbf{u}(\mathbf{x})$ is the deterministic flow, and $d\mathbf{W}(t)$ denotes the forward difference of the vector Wiener process
with statistics \cite{Gardiner_2009}
\begin{equation}
\langle dW_i(t) \rangle = 0, \: \langle dW_k(t) dW_l(t) \rangle =  \delta_{kl} dt.
\label{eq:Wiener}
\end{equation}  
$\sigma$ is a constant \textit{noise tensor} which determines the linear combinations of the $dW_i$ which induce fluctuations in each component of $d\mathbf{x}$.  On the account of (\ref{eq:Wiener}), we think of $d\mathbf{W}$ as formally $O(\sqrt{dt})$.

     The structure of the paper is as follows.  In Sec. II, we examine the projection of statistics in the full Hamiltonian phase space with all the degrees of freedom, including the microscopic, onto a \textit{reduced} statistics which sees only the macroscopic degrees of freedom  \footnote{This also provides helpful insight when we apply metrics such as the stochastic line integral directly to real experimental data collected from high-dimensional systems. The process of experimentally measuring only a few macroscopic variables is qualitatively related to the theoretical process of projecting from a high-dimensional microscopic model (e.g., with number of degrees of freedom of order of Avogadgro's number)  to SDE models described by (\ref{eq:SDE_general}).}. Elementary analysis shows that the projected statistics is characterized by a continuity equation for $\rho (\mathbf{x}, t)$, the marginal of the phase space probability density.  This marginal has an associated probability current $\mathbf{J} (\mathbf{x}, t)$.  There is no \textit{a priori} determination of this ``exact" probability current as a simple functional of the reduced probability density $\rho$.  Nevertheless, we show that \textit{microscopic reversibility} in the sense of Onsager \cite{Onsager_1931} implies that $\mathbf{J}$ vanishes identically.  Conversely, \textit{nonvanishing} of $\mathbf{J}$ indicates irreversibility.   

In Sec. III, we examine line integrals over trajectories on macroscopic state space projected from the full Hamiltonian phase space.
They are useful detectors of irreversibility as pointed out in recent theoretical and experimental work that focused on one particular realization called stochastic area \cite{Ghanta2017, Gonzalez_PRE_2019}.  To develop a general definition, let 
\begin{equation}
C(t): \; \mathbf{x} = \mathbf{x}(t'), \; 0 < t' < t
\end{equation}
be such a trajectory.  Form the line integral
\begin{equation}
G(t) :=\int_{C(t)} g_i \; dx_i = \int_0^t g_i(\mathbf{x}(t')) \; \dot{x}_i(t') \;dt',
\label{eq:line_int}
\end{equation}
where $g_i$ is an arbitrary vector-valued function of $\mathbf{x}$ and a summation convention is used, i.e., repeated Cartesian indices are summed over.  (For the special case of stochastic area, one chooses $g_1 = -\frac{x_2}{2}, \; g_2 = \frac{x_1}{2}$.)
We show that the ensemble average generally satisfies
\begin{equation}
\frac{d}{dt}\left\langle G \right\rangle (t) := \frac{d}{dt}\left\langle \int_{C(t)} g_i \; dx_i \right\rangle =  \int_{\mathbb{R}^n} g_i(\mathbf{x}) J_i(\mathbf{x}, t) d\mathbf{x},
\label{eq:line_int_form}
\end{equation} 
so that the nonvanishing of ensemble averaged line integrals indicates irreversibility.  In the general non-stationary case we expect $d\langle G \rangle /dt$ to have explicit time-dependence.  However, in the important case where probability current $J_i$ is stationary or quasi-stationary, $d\langle G \rangle /dt$ will be time-independent.  

The statistics on the macroscopic state space governed by the stochastic ODE approximates the ``exact" statistics based upon projection from the full Hamiltonian phase space. In Sec. IV, we examine the analog of the stochastic line integral formula (\ref{eq:line_int_form}) within the statistics governed by the SDE (\ref{eq:SDE_general}).  We focus on those systems where the exact probability current $\mathbf{J}$  can be reasonably approximated by a Fokker-Planck probability current \cite{Gardiner_2009},
\begin{equation}
\mathbf{j}(\mathbf{x},t) := \mathbf{u}(\mathbf{x}) \rho - D \nabla \rho.
\label{eq:Fokker_Planck_current}
\end{equation}  
Here, $\rho(\mathbf{x},t)$ denotes the reduced probability density over the macroscopic position variables, and 
\begin{equation}
D := \frac{1}{2}\sigma \sigma^T
\label{eq:Diffusion_tensor}
\end{equation}  
denotes the \textit{diffusion tensor} induced by the noise tensor $\sigma$ in (\ref{eq:SDE_general}).  We propose and analyze two definitions of stochastic line integral closely related to the Stratonovich stochastic integral \citep{vanKampen_2007, Gardiner_2009}, and show their equivalence.  For both, we find that the stochastic line integral formula (\ref{eq:line_int_form}) holds with the Fokker-Planck probability current $\mathbf{j}$ in place of $\mathbf{J}$.

To illustrate the general points in earlier sections and to show connections of stochastic line integrals with physically relevant concepts such as heat transfer, Sec. V examines the overdamped statistics of a well-established paradigm system of two coupled Brownian particles in heat baths of temperatures $T_1$ and $T_2$, cf. Fig. \ref{fig1}.  Heat transfer between the baths is analyzed for the case of linear coupling, starting from its characterization by a specific type of the stochastic line integral, the stochastic area \cite{Ghanta2017}.  For two-dimensional systems, simulations may be complemented by a theoretical quantification of heat transfer in terms of the \textit{stream function} for the stationary probability current.  We show how stochastic line integrals may generally be expressed as relatively simple integrals of the stream function and how this allows to analytically infer physically significant properties such as heat transfer rate. Finally, in Sec. VI we discern the effects of nonlinearity by studying the scaling dependence of time rate of change in stochastic area versus noise amplitude.  In particular, we show analytically that ``hard spring" nonlinearity inhibits heat transfer relative to a linear system, as the temperature difference increases.

\section{Projection of Hamiltonian phase space onto macroscopic variables}
\label{sec:line int}

  To understand the broad applicability of SLIs it is useful to consider how they follow from a microscopic perspective.  Here we focus on high-dimensional closed $classical$ Hamiltonian systems and carry out projections \cite{Zwanzig_2001}. However, the motivation to do this is not purely formal since, for a wide range of experimental noise-driven non-equilibrium systems, one can in princple associate a set of classical Hamiltonians provided there are enough degrees of freedom and the functional form of the Hamiltonian $H$ is appropriately chosen.  Equivalently, we expect that a classical experimental system can be simulated with aribitrary accuracy by a reversible classical computer which in turn can be associated with a classical Hamiltonian flow \cite{Bennett_1982}.  
  
  Let $\mathbf{X}$ and $\mathbf{P}$ be $N$-vectors of coordinates and momenta of the full Hamiltonian dynamics.  We identify the first $n \ll N$ components of $\mathbf{X}$ as the macroscopic coordinates, so the projection of the trajectory $\begin{pmatrix} \mathbf{X}(t) \\ \mathbf{P}(t) \end{pmatrix}$ in $2N$-dimensional phase space onto the space of macroscopic coordinates is 
  \begin{equation}
  \mathbf{x} := \begin{pmatrix} X_1(t) \\ \vdots \\ X_n(t) \end{pmatrix}.
  \nonumber
  \end{equation}
  For convenience of discussion, we denote all of the remaing microscopic coordinates by
  \begin{equation}
  \mathbf{\tilde{X}} := \begin{pmatrix} X_{n+1}(t) \\ \vdots \\ X_N(t) \end{pmatrix}.
  \nonumber
  \end{equation}
  
Figure 2 visualizes the projection from Hamiltonian phase space onto the state space of macroscopic coordinates.  Any ensemble of trajectories on the full phase space is transported by the Hamiltonian flow, so the full probability density $f(\mathbf{X}, \mathbf{P}, t)$ satisfies a continuity condition, equivalently the classical Liouville equation \cite{Zwanzig_2001}
\begin{equation}
\frac{\partial}{\partial t} f + \frac{\partial}{\partial X_i}\left(\frac{\partial H}{\partial P_i} f\right) -  \frac{\partial}{\partial P_i}\left(\frac{\partial H}{\partial X_i} f\right) = 0.
\label{eq:Classical_Liouville}
\end{equation}
Here, $H(\mathbf{X}, \mathbf{P})$ is the (autonomous) Hamiltonian.  Given an ensemble of trajectories on the full phase space, we examine its projection onto the space of macroscopic coordinates.  The probability density of the projected ensemble is
\begin{equation}
\rho (\mathbf{x}, t) := \int_{\mathbb{R}^{N-n}}d\mathbf{\tilde{X}} \; \left( \int_{\mathbb{R}^N} d\mathbf{P} \; f \right) .
\end{equation}
Here, $d\mathbf{\tilde{X}} := d\mathbf{X_{n+1}} \dots d\mathbf{X_N}$ is the volume element of microscopic coordinates.  By integrating (\ref{eq:Classical_Liouville}) over \textit{all} momenta and microscopic coordinates, we arrive at the projection of the continuity equation onto the space of macroscopic coordinates,
\begin{equation}
\frac{\partial}{\partial t} \rho + \sum_{i = 1}^n \frac{\partial}{\partial X_i} \int_{\mathbb{R}^{N - n}}d\mathbf{\tilde{X}} \left(\int_{\mathbb{R}^N}d\mathbf{P}\; \frac{\partial H}{\partial P_i} f \right) = 0.
\end{equation}
Here, we recognize
\begin{equation}
J_i(\mathbf{x}, t) := \int_{\mathbb{R}^{N - n}}d\mathbf{\tilde{X}} \left(\int_{\mathbb{R}^N} d\mathbf{P} \;\frac{\partial H}{\partial P_i} f \right) 
\label{eq:Projected_Prob_Current}
\end{equation}
for $i = 1, \dots n$ as the components of projected probability current in the space of macroscopic coordinates.
\begin{figure}
\centerline{\includegraphics[width=0.40\textwidth]{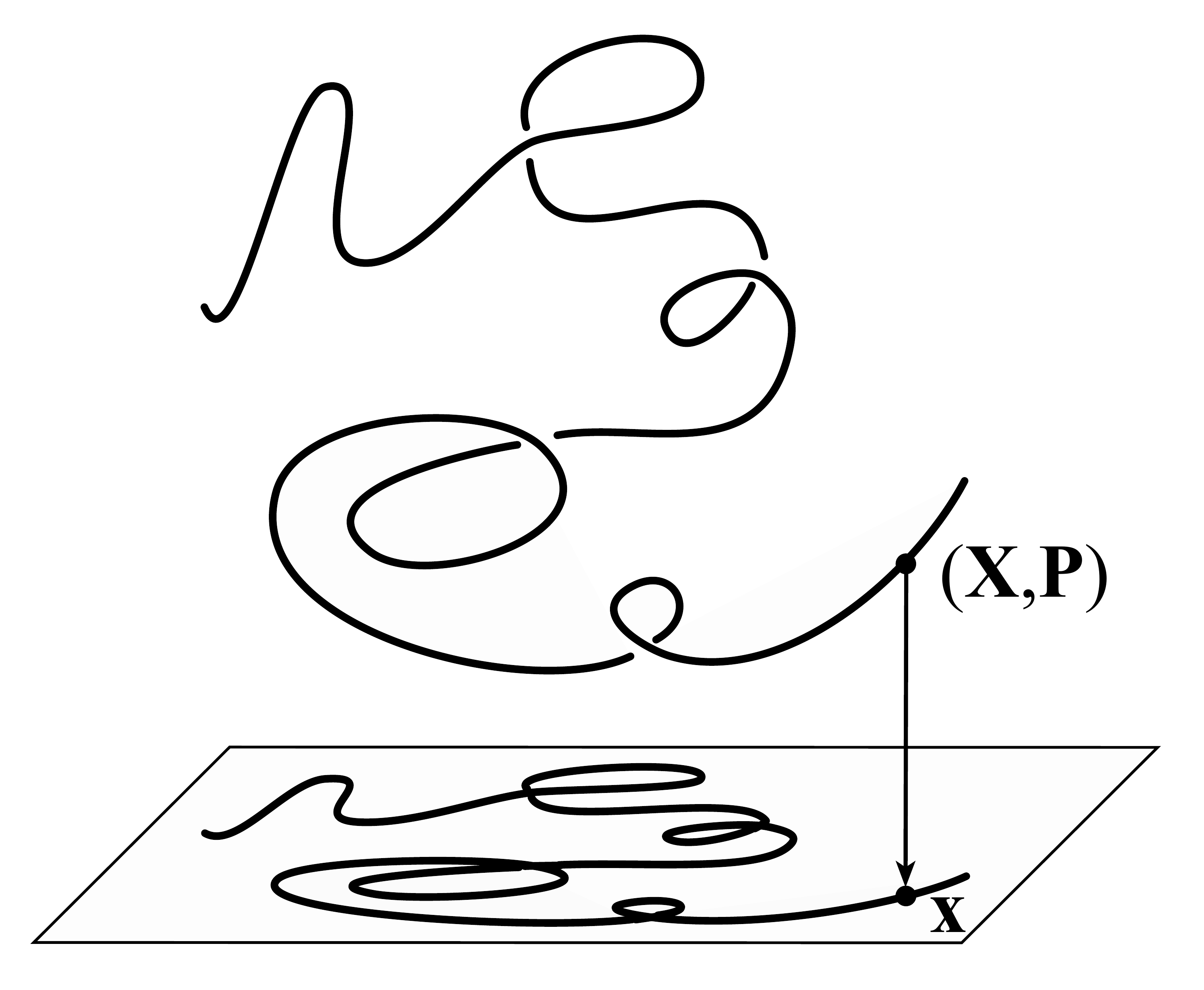}}
\caption{Depiction of the projection from full phase space onto the space of macroscopic states.}
\label{fig2}
\end{figure}

As expected for generic Hamiltonian flows, we assume that $H(\mathbf{X},\mathbf{P})$ is \textit{even} in the momenta \cite{Zwanzig_2001}.  If $\begin{pmatrix} \mathbf{X}(t) \\ \mathbf{P}(t) \end{pmatrix}$ is a realizable phase space trajectory, then so is its \textit{time reversal} $\begin{pmatrix} \mathbf{X}(-t) \\ -\mathbf{P}(-t) \end{pmatrix}$.  Following Onsager \cite{Onsager_1931}, we assume an \textit{equilibrium} ensemble in phase space which is stationary and invariant under interchange of all trajectories with their time reversals.  Let $f_s(\mathbf{X}, \mathbf{P})$ be the stationary probability density of the original ensemble.  At $t = 0$, swap every trajectory for its time reversal.  The probability density after the swap is $f_s(\mathbf{X}, -\mathbf{P})$.  Then, due to the aforementioned invariance, we must have
$f_s(\mathbf{X}, \mathbf{P}) = f_s(\mathbf{X}, -\mathbf{P})$, so that $f_s$ is \textit{even} in $\mathbf{P}$.  Given the even symmetries of $H$ and $f_s$ in the momenta, we see that the projected probability current (\ref{eq:Projected_Prob_Current}) must vanish for all $\mathbf{x}$.  We anticipate that equilibrium modeled by SDE (\ref{eq:SDE_general}) should likewise have zero probability current.  In the context of overdamped SDEs such as (\ref{eq:SDE_general}), this property is called \textit{detailed balance} \cite{Tolman1938, Luchinsky_RPP_1998}.  It is equivalent to \textit{microscopic reversibility} because the forward direction of time cannot be inferred from the stationary reduced statistics. 

Here we are interested in systems where the reduced statistics is \textit{irreversible}, i.e., non-vanishing projected probability current.  What does irreversibility of reduced statistics mean in the context of a much larger whole which is assumed to be reversible and closed?  An illustrative scenario is the situation of two heat baths, each with many degrees of freedom, coupled to one another via a few macroscopic degrees of freedom.  Figure \ref{fig1} visualizes a mechanical example which has been studied extensively as a paradigm non-equilibrium system that breaks detailed balance, see, e.g., \cite{Ciliberto_PRL_2013, Ciliberto_JSM_2013b, Gingrich_2019}.  A mass-spring system consists of two Brownian particles moving in  baths at different temperatures.  They are trapped by restoring forces, and coupled to each other by a linear spring. A temperature difference between the baths drives heat transfer from hot to cold.  Strictly speaking, statistics on the full phase space (including all the microscopic degrees of freedom) cannot be stationary.  Nevertheless, we anticipate that stationary reduced statistics representing the steady heat transfer is achieved asymptotically as the heat capacities of the baths become very large.  This means that the projected statistics of the full Hamiltonian system must settle into a quasi-stationary state for a very long time - longer than the duration of any measurement. Conversely, the lower bound on the time scale for quasi-stationary behavior is limited by the transients to decay from typical arbitrary initial conditions to the quasi-stationary state, a time scale that is, relatively speaking, many of orders of magnitude smaller.  For overdamped systems, this decay time is typically determined by linearizing the deterministic part of the flow in the SDE \footnote{More generally, it less clear that one might infer this separation of time scales looking directly at the structure of the full $H$.  Related to this point, there are certainly many Hamiltonian flows for which this form of separation of time scales is not expected \cite{Zwanzig_2001}.}. 
%
%
%
%
\section{Stochastic Line Integrals in the Full Phase Space}
\label{sec:line int}

In this section, we show generally how stochastic line integrals can be written in terms of the projected probability currents.  Let
\begin{equation}
\frac{d}{dt}\langle G \rangle := \frac{d}{dt} \langle \int_{C(t)} g_i dx_i \rangle  = \langle g_i(\mathbf{x}) \; \dot{x}_i \rangle
\label{eq:Stoch_Line_Int_2}
\end{equation}
be the time rate of change of the ensemble average of the line integral (\ref{eq:line_int}).  Carrying out the ensemble averaging in the full Hamiltonian phase space, we have
\begin{eqnarray}
& & \frac{d}{dt}\langle G \rangle  =  \int_{\mathbb{R}^{2N}} d\mathbf{X} d\mathbf{P}  \; g_i(\mathbf{x}) \frac{\partial H}{\partial P_i}(\mathbf{X}, \mathbf{P}) \; f(\mathbf{X}, \mathbf{P}, t)   = \nonumber \\ & & \int_{\mathbb{R}^n}  d\mathbf{x} \; g_i(\mathbf{x}) \left[ \int_{\mathbb{R}^{N - n}} d\mathbf{\tilde{X}} \int_{\mathbb{R}^N}d\mathbf{P} \; \frac{\partial H}{\partial P_i} f(\mathbf{X}, \mathbf{P}, t) \right]   =  \nonumber \\ & & \int_{\mathbb{R}^n} d\mathbf{x} \; g_i(\mathbf{x}) J_i(\mathbf{x}, t),
\label{eq:Stoch_Line_Int_3}
\end{eqnarray}  
where $J_i(\mathbf{x}, t)$ is the $i$-th component of the projected probability current in (\ref{eq:Projected_Prob_Current}).  From (\ref{eq:Stoch_Line_Int_2}) and (\ref{eq:Stoch_Line_Int_3}),  the previously mentioned line integral formula (\ref{eq:line_int_form}) follows.

We now examine two instructive special cases.  If $\mathbf{g}$ is the differential of a function $\phi(\mathbf{x})$, so $\mathbf{g} = \nabla \phi$, we have
\begin{equation}
g_iJ_i = \partial_i\phi J_i = \nabla \phi \cdot \mathbf{J}.
\end{equation}
For stationary statistics, we have $\nabla \cdot \mathbf{J} = 0$, in which case we can write
\begin{equation}
g_iJ_i = \nabla \cdot (\phi \mathbf{J}). \nonumber
\end{equation}
Hence, the integral (\ref{eq:Stoch_Line_Int_3}) of $g_iJ_i$ over a region of $\mathbb{R}^N$ is a boundary integral.  Assuming sufficiently strong decay of $\phi \mathbf{J}$ as $| \mathbf{x} | \rightarrow \infty$, the boundary integral vanishes as the region expands to the whole $\mathbb{R}^N$. 
Hence, $d\langle G \rangle/ dt$ vanishes for stationary statistics.   We conclude that a line integral of an exact differential form cannot detect irreversibility.

Next, we look at the \textit{stochastic area} \cite{Ghanta2017, Gonzalez_PRE_2019} as a detector of irreversiblity.  Let
\begin{equation}
C(t): x_1 = x_1(t'), \; x_2 = x_2(t'), \; 0 < t' < t,
\end{equation}
be the projection of a stochastic trajectory in time interval $(0, t)$ onto the $x_1, x_2$ plane.  Then, the line integral
\begin{equation}
A(t):= \frac{1}{2}\int_{C(t)} x_1 dx_2 - x_2 dx_1
\label{eq:Area}
\end{equation}
represents the area swept out by the displacement vector $(x_1 , x_2)$ in the time interval $0  < t' < t$.  
Net rotation of the displacement vector about the origin after a sufficiently long time indicates irreversibility.
According to (\ref{eq:Stoch_Line_Int_3}) and (\ref{eq:Area}), the ensemble rate of change of area can be written in terms of $J_i$ as
\begin{equation}
\frac{d}{dt}\langle A \rangle = \frac{1}{2} \int_{\mathbb{R}^n} d\mathbf{x} \; (x_1J_2 - x_2J_1) .
\label{eq:Stoch_Area_Formula}
\end{equation}
  
%
%
%
%
\section{Stochastic line integrals within reduced statistics}
\label{sec:Line integral}

Since the exact statistics of Hamiltonian phase space trajectories projected onto the macroscopic state space is modeled by the SDE (\ref{eq:SDE_general}), it is natural to ask:  what is the analog of the line integral formula (\ref{eq:Stoch_Line_Int_3}) within the reduced statistics?  The analoge of the RHS of (\ref{eq:Stoch_Line_Int_3}) seems clear enough:  the exact probability current $\mathbf{J}$ can be replaced by the Fokker-Planck probability current, cf. (\ref{eq:Fokker_Planck_current}).  A deeper question concerns how to implement line integrals over trajectories of the SDE (\ref{eq:SDE_general}).   Here, we examine two definitions of stochastic line integral closely both of which are related to the Stratonovich stochastic integral \cite{Gardiner_2009}.

The first proposed definition of stochastic line integral is
\begin{equation}
\int_{C(t)} g_i(\mathbf{x}) dx_i := \lim_{N \rightarrow \infty} \sum_{t'} g(\mathbf{y}(t')) dx_i(t').
\label{eq:SLI_Stratonovich}
\end{equation}
Here, $t' := ndt$, $dt := t/N$, and the summation runs from $n = 0$ to $n = N$.  The 
$dx_i(t')$ are 
components of the forward difference as in (\ref{eq:forward_diff}), and the $\mathbf{y}(t')$ are \textit{midpoints}
\begin{equation}
\mathbf{y}((t')) := \frac{\mathbf{x}(t') +  \mathbf{x}(t '+ dt)}{2}.
\end{equation}
The evaluation of differential form components $g_i$ at midpoints between $\mathbf{x}(t')$ and $\mathbf{x}(t' + dt)$ is analogous to the 
Stratonovich stochastic integral \cite{Gardiner_2009, vanKampen_2007}.

The second definition of stochastic line integral replaces evaluation of the $g_i$ at \textit{midpoints} by the average at endpoints.  that is:
\begin{eqnarray}
& & \int_{C(t)} g_i(\mathbf{x}) dx_i := \nonumber\\  & & \lim_{N \rightarrow \infty} \sum_{t'} \frac{1}{2} \left[ g_i(\mathbf{x}(t')) + g_i(\mathbf{x}(t' + dt)) \right] dx_i(t').
\label{eq:SLI_midpoints}
\end{eqnarray}
Here, the summation is over $t' = n dt, \: 0 \leq n < N$ as in the first defintion (\ref{eq:SLI_Stratonovich}).  Let's look at the difference between the two definitions,
\begin{equation}
\lim_{N \rightarrow \infty} \sum_{t'} \left[ \frac{1}{2}g_i(\mathbf{x}(t')) - g_i(\mathbf{y}(t')) + \frac{1}{2}g_i(\mathbf{x}(t' + dt)) \right] dx_i(t').
\label{eq:SLI_difference}
\end{equation}
Setting $\mathbf{x}(t') = \mathbf{y}(t') - \frac{1}{2}d\mathbf{x}(t'), \: \mathbf{x}(t' + dt) = \mathbf{y}(t') + \frac{1}{2}d\mathbf{x}(t')$, we find that the Talyor polynomial approximation of the term in brackets is quadratic in $d\mathbf{x}(t')$, so is $O(dt)$. Then the entire sum in (\ref{eq:SLI_difference}) is $O(N dt^{\frac{3}{2}}) = O(\frac{1}{\sqrt{N}}) \rightarrow 0$ as $N \rightarrow \infty$. 
The limit process of the second definition yields the same result as the first.  Why the second definition? Its equivalence to the first gives a reassuring sense that the notion of stochastic line integral is robust.  In addition, the two definitions of stochastic line integral lead to two constructions of probability current based upon discrete time series of state variables, and close variants of both appear in the literature \cite{Gingrich_2019,Gonzalez_PRE_2019}.

Our immediate task is to analyze the ensemble-averaged stochastic line integral according to the first definition (\ref{eq:SLI_Stratonovich}). We take
 the time series $\mathbf{x}(t)$ to be a solution of the \^{I}to SDE (\ref{eq:SDE_general}). Setting 
$\mathbf{y}(t') = \mathbf{x}(t') + \frac{1}{2} \mathbf{dx}(t')$, and 
evoking (\ref{eq:forward_diff}) for the forward difference $\mathbf{dx}(t')$, we have
\begin{eqnarray}
& &\langle g_i(\mathbf{y}(t')) dx_i(t') \rangle = \nonumber \\ & & \langle (g_i(\mathbf{x}(t')) 
dx_i(t') \rangle + \nonumber \\  & & \frac{1}{2}\langle (\partial_j g_i)(\mathbf{x}(t')) dx_i(t') dx_j(t') 
\rangle + O(dt^{\frac{3}{2}}) = \nonumber \\  & &  \langle g_i (\mathbf{x}(t'))\sigma_{ij} 
dW_j(t') \rangle + \langle (g_i u_i )(\mathbf{x}(t') \rangle dt +  \nonumber \\  & & \frac{1}{2} 
\langle (\partial_j g_i)(\mathbf{x}(t')) \sigma_{ik} \sigma_{jl} dW_k(t') dW_l(t') \rangle + 
O(dt^{\frac{3}{2}}).
\end{eqnarray}

Since $\mathbf{x}(t')$ is independent of the increment $d\mathbf{W}(t')$ over time interval $(t', t' + dt)$, we simplify the RHS to
\begin{eqnarray}
& &\langle g_i (\mathbf{x}(t')) \rangle  \sigma_{ij} \langle dW_j(t') \rangle + \langle (g_i u_i )(\mathbf{x}(t') \rangle \tau + \nonumber\\  & & \frac{1}{2} \langle (\partial_j g_i)(\mathbf{x}(t')) \rangle  \sigma_{ik} \sigma_{jl} \langle dW_k dW_l(t') \rangle + O(dt^{\frac{3}{2}}).
\end{eqnarray}
Evoking the statistics of $dW_i$, cf. (\ref{eq:Wiener}), we can write
\begin{equation}
\langle g_i(\mathbf{y}(t')) dx_i(t') \rangle = \langle (g_i u_i + D_{ij} \partial_j g_i )(\mathbf{x}(t')) \rangle dt + O(dt^{\frac{3}{2}}).
\end{equation}
Here, $D_{ij}$ are the components of the diffusion tensor in (\ref{eq:Diffusion_tensor}).  Hence,
\begin{eqnarray}
& & \langle \sum g_i(\mathbf{y}(t')) dx_i(t') \rangle =  \nonumber\\ & & \sum\langle (g_i u_i + D_{ij} \partial_j g_i )(\mathbf{x}(t')) \rangle dt + O(\frac{1}{\sqrt{N}}).
\end{eqnarray}
The RHS is the Riemann sum for the integral of $\langle (g_i u_i + D_{ij} \partial_j g_i )(\mathbf{x}(t') \rangle$ over $0 < t' < t$, so we have
\begin{equation}
\left\langle \int_{C(t)} g_i(\mathbf{x}) dx_i \right\rangle = \int_0^t dt' \left\langle (g_i u_i + D_{ij} \partial_j g_i )(\mathbf{x}(t')) \right\rangle .
\label{eq:Line_integral_FP_expression}
\end{equation}
Introducing the reduced probability density $\rho (\mathbf{x}, t)$, we have
\begin{eqnarray}
 \left\langle \left(g_i u_i + D_{ij} \partial_j g_i \right) \right\rangle =  \nonumber \\ \int_{\mathbb{R}^n} \mathbf{dx} \rho (g_i u_i + D_{ij} \partial_j g_i )   = \nonumber \\ \int_{\mathbb{R}^n} \mathbf{dx} g_i (\rho u_i - D_{ij} \partial_j \rho ) .
\label{eq:FP_average}
\end{eqnarray}
Here, we recognize the components of the Fokker-Planck probability current (\ref{eq:Fokker_Planck_current}), equivalently the reduced probability current density.  Thus,  (\ref{eq:Line_integral_FP_expression}) and (\ref{eq:FP_average}) allow us to write the line integral formula within the reduced statistics,
\begin{equation}
\frac{d}{dt}\langle G \rangle  = \int_{\mathbb{R}^n} d\mathbf{x} g_i j_i .
\label{eq:Stoch_line_int_reduced}
\end{equation}
This equation has the same form as (\ref{eq:Stoch_Line_Int_3}) with projected probability current $\mathbf{J}$ now replaced by Fokker-Planck current $\mathbf{j}$, and reaffirms the role of the stochastic line integrals as a detector of irreversibility.  For steady-state situations (i.e., $\partial\mathbf{j}/\partial t = 0$), if we have $\mathbf{j} = 0$ everywhere in the reduced phase space, then $d\langle G \rangle /dt$ vanishes for any choice of vector function $\mathbf{g}$ and detailed balance is satisfied.  Conversely, if $\mathbf{j}$ has regions of non-zero value, one can choose $\mathbf{g}$ such that $d\langle G \rangle /dt \not= 0$ with the magnitude and sign determined according to (\ref{eq:Stoch_line_int_reduced}), thereby providing a quantitative measure of the associated irreversible behavior.  Suitable choices of $\mathbf{g}$ will typically depend on system details; examples are discussed in the following section.


%
%
%
%

%
%
%
%
\section{Irreversible statistics of coupled Brownian particles -  stream function approach}
\label{sec:Brownian particles}

We analyze the overdamped statistics of two degrees of freedom with coordinates $x_1$ and $x_2$ in a given energy landscape $U(x_1, x_2)$.  Two major reasons for studying this type of system include: 1) the physics is analogous to that of many physical non-equilibrium systems that are the focus of recent work including coupled electronic circuits and mechanical systems \cite{Gonzalez_PRE_2019, Chiang_2017a, Chiang_2017b, Gnesotto_2018}; and 2) the reduction of high-dimensional systems onto two-dimensional planar subspaces often has a similar effective dynamics \cite{Battle_Science_2016, Ciliberto_PRL_2013}.  The \^{I}to SDE system for $x_1(t)$ and $x_2(t)$ can be written as
\begin{equation}
dx_1 = -\mu_ 1 \partial_1 U dt  + \sigma_1 dW_1, \: \: dx_2 = -\mu_2 \partial_2 U dt + 
\sigma_2 dW_2,
\label{eq:SDE_2d_model}
\end{equation}
where $\mu_1$ and $\mu_2$ are given \textit{mobilities}, $dW_1$ and $dW_2$ are 
forward differences of independent Wiener processes, and  $\sigma_1$ and $\sigma_2$ are given noise amplitudes.  Following previous studies \citep{Ciliberto_JSM_2013b, Chiang_2017b, Gingrich_2019}, we can think of (\ref{eq:SDE_2d_model}) as the stochastic dynamics of two Brownian particles in heat baths of temperatures $T_1$ and $T_2$.  Each particle moves in its own potential, which may be nonlinear, and we assume below that they are coupled to one another with a linear (or nonlinear) spring, cf. Fig. \ref{fig1}.  The noise amplitudes are related to the temperatures by the usual Einstein relations \cite{vanKampen_2007}
\begin{equation}
\sigma_1 = \sqrt{2\mu_1 k_B T_1}, \: \:\sigma_2 = \sqrt{2\mu_2 k_B T_2}.
\label{eq:sigma12_defn}
\end{equation}

We now focus on the specific form taken by the line integral formula (\ref{eq:Stoch_line_int_reduced}) for stationary statistics of the two-dimensional stochastic dynamics of (\ref{eq:SDE_2d_model}).  The stationary probability current $\mathbf{j}$ on $\mathbb{R}^2$ is divergence free, so there exists a \textit{stream function} $\psi(\mathbf{x})$ satisfying \cite{Neu_Training_2009}
\begin{equation}
j_1 = \partial_2 \psi, \; j_2 = -\partial_1 \psi.
\label{eq:Stoch_stream_defn}
\end{equation}
Assuming that $\mathbf{j}$ decays to zero sufficiently rapidly as $|\mathbf{x}| \rightarrow \infty$, we may select the unique stream function which vanishes at infinity.  The reason for this selection becomes evident momentarily: substituting (\ref{eq:Stoch_stream_defn}) for $j_1, j_2$ into the line integral formula (\ref{eq:Stoch_line_int_reduced}), we have
\begin{equation}
\frac{d}{dt}\langle G \rangle:= \int_{\mathbb{R}^2} d\mathbf{x} (g_1 \partial_2 \psi - g_2 \partial_1 \psi)  = \int_{\mathbb{R}^2} d\mathbf{x} \psi ( \partial_1 g_2 - \partial_2 g_1),
\end{equation}
where the second equality uses integration by parts and the boundary condition that $\psi$ vanishes at $\infty$.  For stochastic area (\ref{eq:Stoch_Area_Formula})  ($g_1 = -\frac{x_2}{2}, \; g_2 = \frac{x_1}{2}$), the ensemble averaged rate of change can be written concisely in terms of the stream function
\begin{equation}
\frac{d}{dt}\langle A \rangle = \int_{\mathbb{R}^2} d\mathbf{x} \psi.
\label{eq:Stream_stoch_area}
\end{equation}
This remarkable formula shows a simple and direct connection between the stochastic area and the stream function.  For some problems it may be relatively easy to compute or even calculate $\psi$ and (\ref{eq:Stream_stoch_area}) provides a direct connection to ensemble-averaged time rate of change of area thus providing a quantitative measure of system irreversibility \cite{Gonzalez_PRE_2019, Gingrich_2019}.

We now formulate a boundary value problem for the stream function, noting that equations of this type have been studied extensively in hydrodynamics, with a large arsenal of solution methods available \cite{Neu_Training_2009, Neu_Singular_2015}.  Recalling that $j_i = -\mu_i 
\partial_i U \rho - \frac{1}{2} \sigma_i^2 \partial_i \rho$, we can recast the stream function derivatives in (\ref{eq:Stoch_stream_defn}) as
\begin{equation}
\partial_2 \psi = -\mu_1 \rho \partial_1 U  - \frac{1}{2} \sigma_1^2  \partial_1 \rho, \; \partial_1 \psi = \mu_2 \rho \partial_2 U  + \frac{1}{2} \sigma_2^2 \partial_2 \rho.
\label{eq:}
\end{equation}
Now, use these expressions to evaluate the combination $-\sigma_1^2 \partial_1 (\partial_1 \psi/\rho) - \sigma_2^2 \partial_2 (\partial_2 \psi/\rho)$ to arrive at
\begin{equation}
-\sigma_1^2 \partial_1 (\frac{\partial_1 \psi}{\rho}) - \sigma_2^2 \partial_2 (\frac{\partial_2 \psi}{\rho}) = (\sigma_2^2 \mu_1 - \sigma_1^2 \mu_2) \partial_{12} U.
\label{eq:Stream_BVP1}
\end{equation}
With the help of (\ref{eq:sigma12_defn}), the prefactor $\sigma_2^2 \mu_2 - \sigma_2^2 \mu_1$ on the RHS of (\ref{eq:Stream_BVP1}) is $2\mu_1 \mu_2 (T_2 - T_1)$ so that the stream function $\psi(\bf{x})$ satisfies 
\begin{equation}
-\sigma_1^2 \partial_1 (\frac{\partial_1 \psi}{\rho}) - \sigma_2^2 \partial_2 (\frac{\partial_2 \psi}{\rho}) = 2\mu_1 \mu_2 (T_2 - T_1) \partial_{12} U,
\label{eq:Stream_BVP2}
\end{equation} 
an inhomogeneous elliptical PDE.  The operator
\begin{equation}
-\sigma_1^2 \partial_1 (\frac{\partial_1}{\rho}) - \sigma_2^2 \partial_2 (\frac{\partial_2}{\rho})
\label{eq:stream_operator}
\end{equation}
on the LHS is positive, i.e.,
\begin{eqnarray}
 \int_{\mathbb{R}^2} d\mathbf{x} \varphi \{-\sigma_1^2 \partial_1 (\frac{\partial_1 \varphi}{\rho}) - \sigma_2^2 \partial_2 (\frac{\partial_2 \varphi}{\rho})\}  &=& \\ \nonumber \int_{\mathbb{R}^2} d\mathbf{x} \{\sigma_1^2 \frac{(\partial_1 \varphi)^2}{\rho} + \sigma_2^2 \frac{(\partial_2 \varphi)^2}{\rho}\} &>& 0,
\end{eqnarray}
for all reasonably behaved $\varphi(\mathbf{x})$ that satisfy zero boundary conditions and are not identically constant.  Hence, the solution to (\ref{eq:Stream_BVP2}) subject to $\psi$ vanishing at infinity is unique.

If the degrees of freedom are uncoupled $(\partial_{12} U = 0)$ or the temperatures are equal $(T_1 = T_2)$, then the stream function and hence probability current vanish identically, cf. (\ref{eq:Stream_BVP2}).  Equivalently, the time rate of change of ensemble averaged line integrals vanishes, cf. (\ref{eq:Stream_stoch_area}). 

We now examine the prototypical example of linearly coupled degrees of freedom with energy landscape
\begin{equation}
U(x_1, x_2) = u_1(x_1) + u_2(x_2) + \frac{k}{2}(x_2 - x_1)^2.
\label{eq:potential_energy}
\end{equation}
Here we treat $u_1(x_1)$ and $u_2(x_2)$ as potential energies of restoring forces acting on each degree of freedom separately.  The two degrees of freedom are coupled by a spring with stiffness $k > 0$.  We assume that $u_1 (x)$ and $u_2 (x)$ diverge to $+\infty$ as 
$|x| \rightarrow \infty$ sufficiently rapidly so that there is stationary probability density satisfying the usual normaliztion condition  $\int_{\mathbb{R}^2} \rho \mathbf{dx} = 1$.  For the energy (\ref{eq:potential_energy}), we have $\partial_{12} U  = -k < 0$, and then the RHS of (\ref{eq:Stream_BVP2}) is a positive constant for $T_1 > T_2$.  Due to the positivity of the operator (\ref{eq:stream_operator}) and utilizing the maximum principle for elliptic partial differential equations \cite{Evans_PDE_2010}, we conclude that $\psi > 0$ for $T_1 > T_2$.  Then the rate of change of stochastic area in (\ref{eq:Stream_stoch_area}) is positive.  This is consistent with the circulation of probability current:  The stream function has a global \textit{maximum}, and the circulation of probability current sufficiently close to this maximum is \textit{counterclockwise}.  A similar argument for the case $T_1 < T_2$ results in $d\langle A \rangle/dt < 0$ and \textit{clockwise} circulation near the global \textit{minimum} of $\psi$.   

Next, we show that the rate of heat transfer between baths is the rate of change of a similar stochastic line integral.  To see this, note that the forward difference of the energy 
(\ref{eq:potential_energy}) is
\begin{equation}
\begin{split}
dU = \partial_1 U dx_1 + \partial _2 U dx_2 + \frac{1}{2} \partial_{11} U (dx_1)^2 + \\
\partial_{12} U dx_1 dx_2  + \partial_{22} U (dx_2)^2 + O(dt^{\frac{3}{2}} ).
\label{eq:potential_energy_t_deriv}
\end{split}
\end{equation}

Substituting for $dx_1$ and $dx_2$ according to the \^{I}to SDE (\ref{eq:SDE_2d_model}), and using the identities (\ref{eq:Wiener}) satisfied by $dW_1$ and $dW_2$,
the ensemble average of (\ref{eq:potential_energy_t_deriv}) yields
\begin{equation}
\frac{d}{dt}\langle U \rangle = -\dot{Q}_1 - \dot{Q}_2,
\end{equation}
where 
\begin{eqnarray}
\dot{Q}_1 &:=& \mu_1 \langle (\partial_1 U)^2 \rangle - \frac{\sigma_1^2}{2} \langle \partial_{11} U \rangle,
\label{eq:Qdot_1}  \\
\dot{Q}_2 &:=& \mu_2 \langle (\partial_2 U)^2 \rangle - \frac{\sigma_2^2}{2} \langle \partial_{22} U \rangle.
\label{eq:Qdot_2}
\end{eqnarray}
For stationary statistics, $\langle U \rangle$ is time independent, so $\dot{Q}_1$ and $-\dot{Q}_2$ have a common value, which 
we denote by $\dot{Q}$.  In the formula (\ref{eq:Qdot_1}) for $\dot{Q}_1$, we interpret $ \mu_1 \langle (\partial_1 U)^2 \rangle$ as the rate of heating of Bath 1 by 
dissipation, and $\frac{1}{2} \sigma_1^2 \langle \partial_{11} U \rangle$ as the rate of work done on the mass by Bath 
1 induced fluctuations.  Hence, $\dot{Q}$ is the net rate of energy transfer 
into Bath 1 and we can express $\dot{Q}$ is the rate of change of the ensemble average of the following stochastic line integral
\begin{equation}
q(t) := -\frac{1}{2} \int_{C(t)} (\partial_1 U dx_1 - \partial_2 U  dx_2).
\end{equation}
Carrying out the ensemble and using (\ref{eq:Stoch_line_int_reduced}) with $g_1 = -\frac{1}{2} \partial_1 U, \: g_2 = \frac{1}{2} \partial_2 U$, we have
\begin{equation}
\frac{d}{dt} \langle q \rangle = -\frac{1}{2} \int_{\mathbb{R}^2} (\partial_1 U j_1 - \partial_2 U j_2 ) \mathbf{dx}.
\label{eq:heat_line_integral}
\end{equation}
Substituting
\begin{equation}
j_1 = -\mu_1 \partial_1 U \rho - \frac{ \sigma_1^2}{2} \partial_1 \rho, 
\: j_2 = -\mu_2 \partial_2 U \rho - \frac{\sigma_2^2}{2} \partial_1 \rho, \nonumber
\end{equation}
we readily find
\begin{equation}
\frac{d}{dt} \langle q \rangle = \frac{1}{2} (\dot{Q}_1 - \dot{Q}_2 ) = \dot{Q}.
\end{equation}
Alternatively, substitute $j_1 = \partial_2 \psi, \: j_2 = -\partial_1 \psi$ into (\ref{eq:heat_line_integral}), to find
\begin{equation}
 \dot{Q} = \frac{d}{dt} \langle q \rangle = \int_{\mathbb{R}^2} d\mathbf{x} \psi \partial_{12} U ,
\label{eq:nonlinear_g_example}
\end{equation}
and we see that there is \textit{no} heat transfer if the degrees of freedom are uncoupled, or the temperatures of the baths are equal.  For \textit{linear} coupling with $\partial_{12} U  := -k < 0$, (\ref{eq:nonlinear_g_example}) reduces to
\begin{equation}
\dot{Q} = - k\frac{d}{dt}\langle A \rangle,
\label{eq:Qdot_formula}
\end{equation}
where $d\langle A \rangle/dt$ is the rate of change of ensemble-averaged stochastic area.  For $T_1 > T_2$ we have $d\langle A \rangle/dt > 0$ and $\dot{Q} < 0$, and for $T_1 < T_2, \; \dot{Q} < 0$, as expected.  It is interesting to note that expression (\ref{eq:nonlinear_g_example}) also applies to the case of nonlinear coupling spring.  For a variety of different nonlinear couplings (e.g., cubic, quartic, Lennard-Jones, etc.) there will be a correponding set of stochastic line integrals that utilize distinct vector $\bold{g}$ functions.

In principle, $d\langle A \rangle /dt$ and hence $\dot{Q}$ are functions of the noise amplitudes $\sigma_1$ and $\sigma_2$, or equivalently, the bath temperatures $T_1$ and $T_2$.  We first consider \textit{linear} stochastic dynamics, with 
\begin{equation}
u_1(x) = u_2(x) = \frac{K}{2} x^2.
\end{equation}
From the theory of linear stochastic dynamics worked out in \cite{Ghanta2017}, we readily find
\begin{equation}
\frac{d}{dt}\langle A \rangle = \frac{k^2}{k + K} \frac{\mu_2 \sigma_1^2 - \mu_1 \sigma_2^2}{\mu_1 + \mu_2}.
\end{equation}
Using (\ref{eq:sigma12_defn}), we can then express $d\langle A \rangle /dt$ in terms of temperatures as

\begin{equation}
\frac{d}{dt}\langle A \rangle = \frac{\mu_1 \mu_2}{\mu_1 + \mu_2} \frac{k}{K + k} k_B (T_1 - T_2).
\label{eq:DADT_dim}
\end{equation}
Now use (\ref{eq:Qdot_formula}) to write heat transfer rate as
\begin{equation}
\dot{Q} = \frac{\mu_1 \mu_2}{\mu_1 + \mu_2} \frac{k^2}{K + k} k_B (T_1 - T_2),
\end{equation}
and it is seen that the rate of heat transfer from Bath 1 to Bath 2 is simply proportional to the temperature difference $T_2 - T_1$.  

We test (\ref{eq:DADT_dim}) by direct numerical solution of the SDE  for the spring-coupled masses (\ref{eq:SDE_2d_model}).  For numerical work, we use dimensionless units and we take $\mu_1 = \mu_2 = \mu$ (common value).  We assume $x_1, \; x_2$ are dimensionless, and we let $E$ denote a characteristic energy associated with $U(x_1, x_2)$.  The unit of time is $(\mu E)^{-1}$, and noise amplitudes are measured in units of $\sqrt{\mu E}$.  Working with dimensionless units, we set $\mu_1 = \mu_2 = 1$ in the SDE (\ref{eq:SDE_2d_model}).  Then, the dimensionless energy (cf. (\ref{eq:potential_energy})) for linear stochastic dynamics is 
\begin{equation}
U(x_1, x_2) = \frac{1}{2} x_1^2 + \frac{1}{2} x_2^2 + \frac{\epsilon}{2} (x_2 - x_1)^2,
\end{equation}
where $\epsilon := \frac{k}{K}$.  We can also write (\ref{eq:DADT_dim}) for $d\langle A \rangle / dt$ in dimensionless form as 
\begin{equation}
\frac{d}{dt}\langle A \rangle = \frac{1}{2} \frac{\epsilon^2}{1 + \epsilon} (\sigma_1^2 - \sigma_2^2).
\label{eq:DADT_dimless}
\end{equation}

Now, let $x_1[t]$ and $x_2[t]$ be discrete time series of $x_1$ and $x_2$ obtained by standard Euler-Maruyama iterations with time step $\tau$, starting from zero initial conditions.  The discrete Stratonovich approximation to the stochastic area at time $t = N \tau$, $N$ a positive integer, is
\begin{equation}
A(t) = \sum_{n = 0}^{N - 1} dA(n \tau),
\end{equation}
where \begin{equation}
dA(t) := \frac{1}{2 \tau} (x_1(t) x_2(t + \tau) - x_1(t + \tau) x_2(t)).
\end{equation}
\begin{figure}
\centerline{\includegraphics[width=0.35\textwidth]{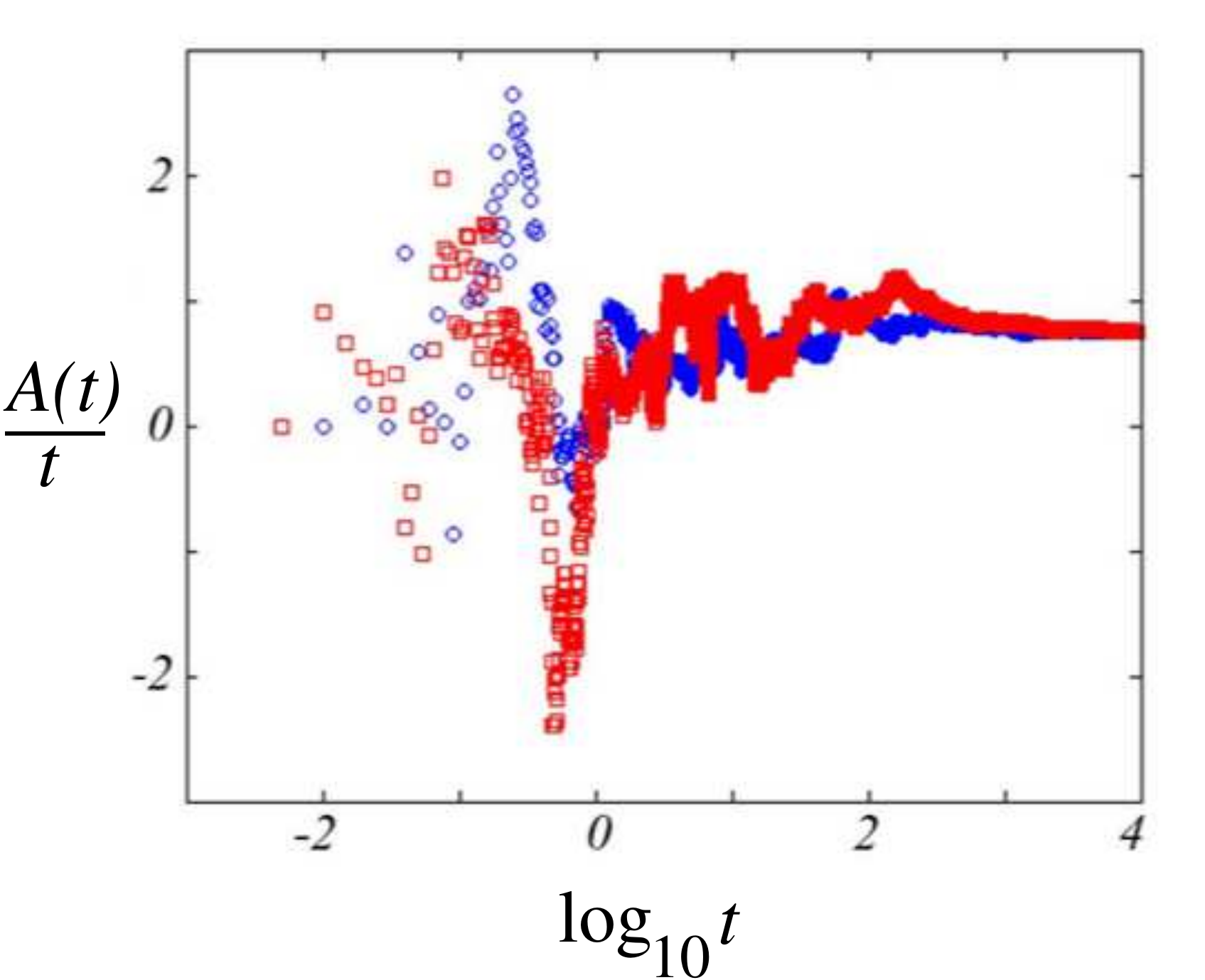}}
\caption{Convergence of $A(t)/t$ for two independent stochastic trials showing approach to a common value associated with the stationary state.}
\label{fig4}
\end{figure}

Figure \ref{fig4} presents numerical results for $A(t)/t$ for two independent trials associated with distinct realizations of the noisy term but with common statistics.  The parameter values are $\epsilon = 1.0$, $\sigma_1 = 2.0$, and $\sigma_2 = 1.0$.  We perform two independent runs, the first of which has $10^6$ time steps with step size $\tau = 0.01$.  The second, with twice as many time steps and half the step size, provides a straightforward confirmation of numerical convergence.  Given a numerical time series of $A(t)$, we plot $A(t)/t$ versus elapsed time $t$ (logarithmic scale).  The (blue) circles are obtained from the first run with $10^6$ time steps.  In the limit of large times, we observe convergence to a constant value, $A(t)/t \approx 0.747$ for $t >> 1$.  The (red) squares are obtained from the second run with $2 \times 10^6$ time steps.  The initial transient is different, since noise forcings for different simulations are independent.  The second run gives  $A(t)/t \approx 0.750$ for $t >>1$, consistent with the value from the first run. Furthermore, both values are consistent with the theoretical prediction based on (\ref{eq:DADT_dimless}) which evaluates to $d\langle A \rangle/dt = 3/4$ for the indicated parameters.

\begin{figure}
\centerline{\includegraphics[width=0.35\textwidth]{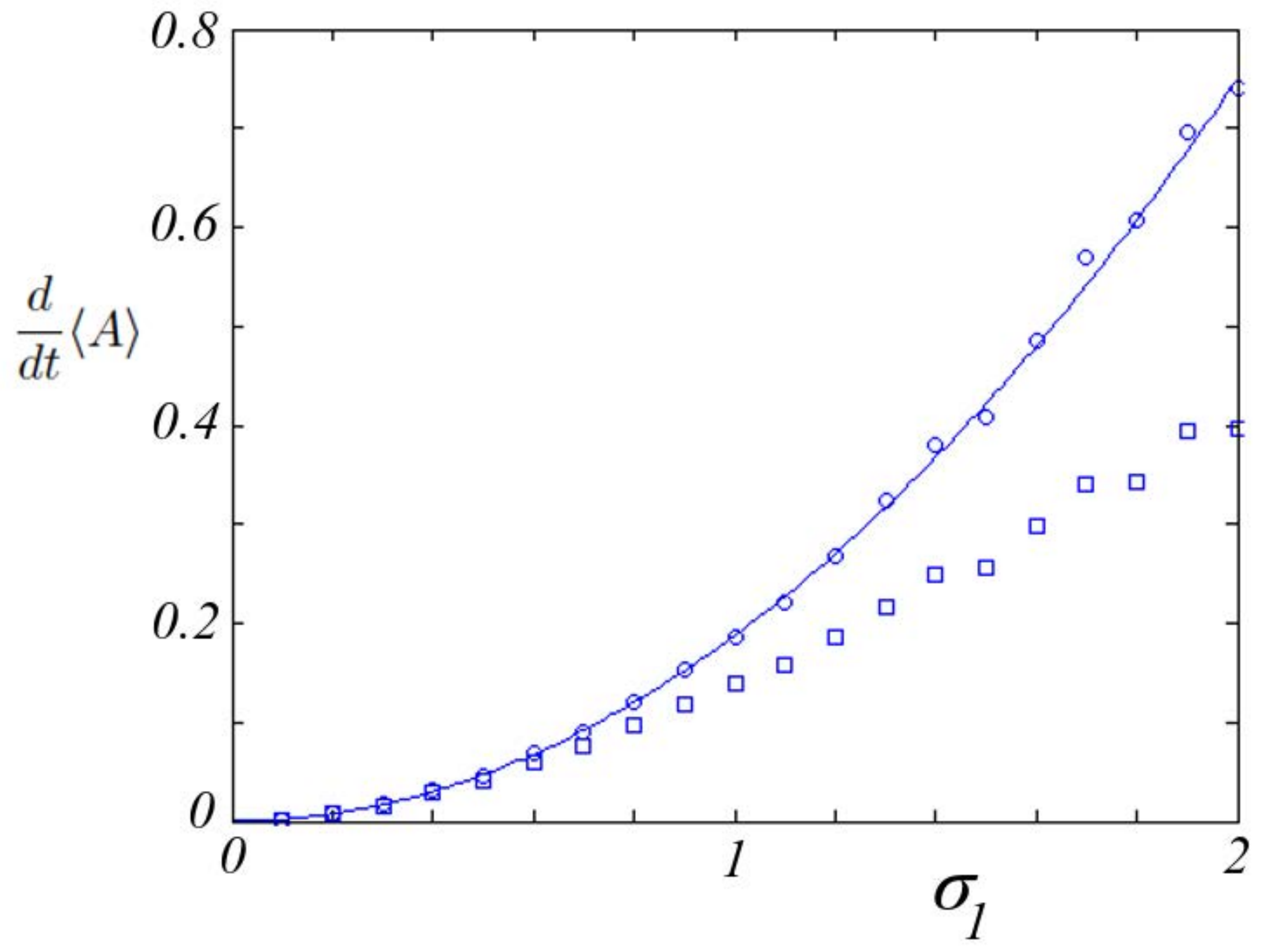}}
\caption{Dependence of time rate of change of stochastic area on noise amplitude for linear (circles) and nonlinear (boxes) restoring forces.  In the linear case, also shown is the theoretical curve obtained by evaluation of (\ref{eq:DADT_dimless}).}
\label{fig5}
\end{figure}

Figure \ref{fig5} displays the dependence of $d\langle A \rangle/dt$ upon the noise amplitude $\sigma:= \sigma_1$ with fixed ratio $r:=\sigma_2/\sigma_1 = 0.5$.  Numerical calculations are carried out for a sequence of $\sigma_1$ values in the range $0$ to $2.0$.  The circles represent estimates of $d\langle A \rangle/dt$ based upon computed values of $A(t)/t$ for elapsed time $t = 10^4$.  The solid curve represents the theoretical dependence of $d\langle A \rangle/dt$ upon $\sigma_1$ with $\sigma_2/\sigma_1 = 0.5$ evaluated according to (\ref{eq:DADT_dimless}).  Numerical data are clearly consistent with quadratic scaling that is predicted by (\ref{eq:DADT_dimless}).  In contrast, the square data points associated with the lower curve are obtained with a nonlinear restoring force and discussed in the following section.

%
%
%
%
\section{Effects of nonlinearity - scaling of stochastic area with noise intensity}
\label{sec:Nonlinearity}

\begin{figure}
\centerline{\includegraphics[width=0.35\textwidth]{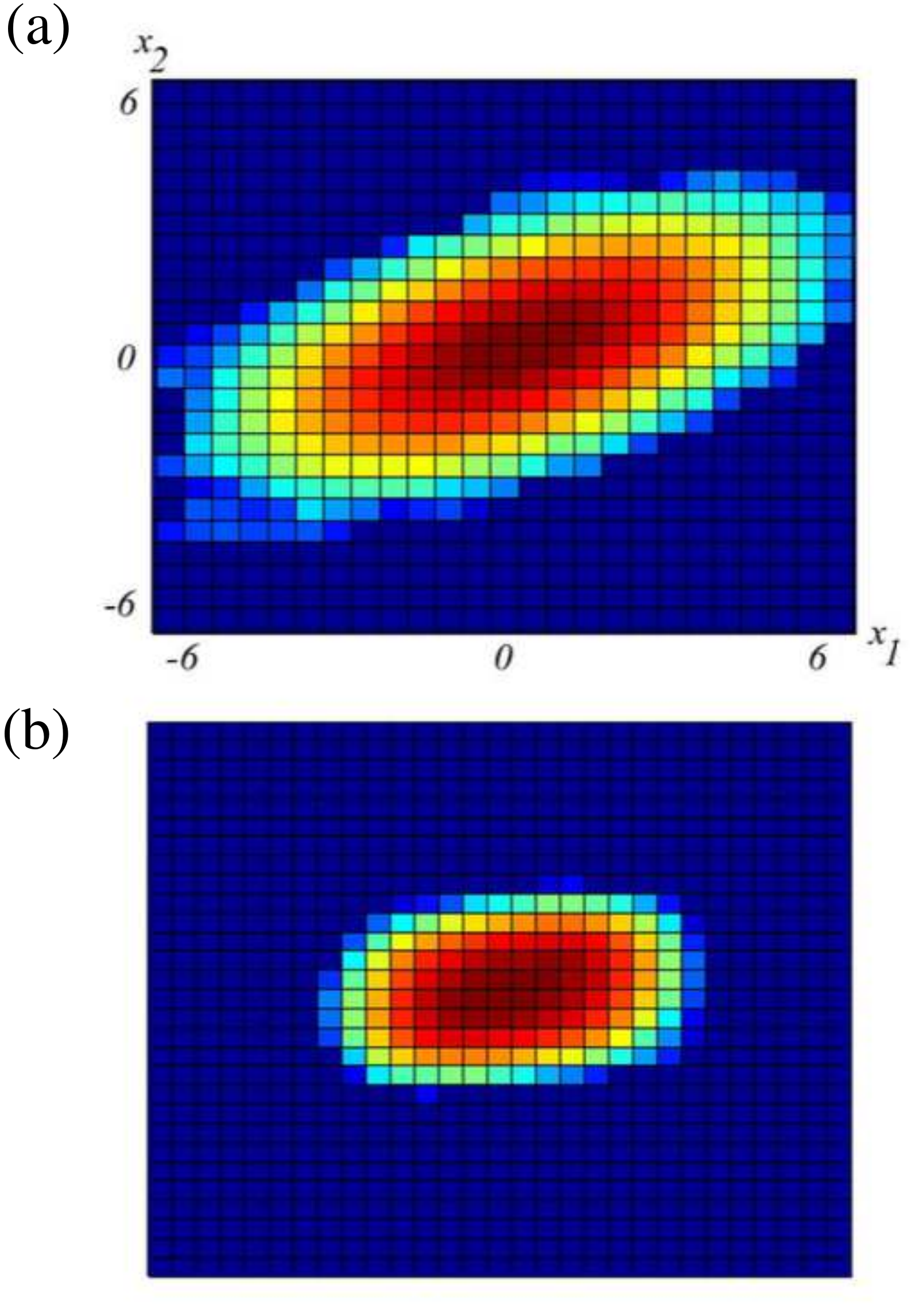}}
\caption{Stationary probability density histograms for systems driven by identical noise amplitudes: (a) linear restoring force, and (b) nonlinear restoring force.}
\label{fig6}
\end{figure}

The dependence of $d\langle A \rangle/dt$ upon noise amplitudes is modified by nonlinearity.  Related to this, the square data points in Fig. \ref{fig5} represent numerical values of $d\langle A \rangle/dt$ obtained by using a potential energy function
\begin{equation}
U(x_1, x_2) = u(x_1) + u(x_2) + \frac{\epsilon}{2} (x_2 - x_1)^2,
\label{eq:potential_energy_nonlin}
\end{equation}
with $u(x)$ is given by 
\begin{equation}
u(x) = \frac{1}{2} x^2 + \frac{1}{4} x^4.
\label{eq:potential_energy_nonlin1}
\end{equation}
This corresponds to a \textit{stiff} spring - at small vibrational amplitude the restoring force is linear while at larger amplitudes the positive quartic term is associated with a increasing effective spring constant.  As $\sigma$ increases from $0$ to $2.0$, the dependence of $d\langle A \rangle/dt$ upon $\sigma$ transitions from quadratic to approximately linear a trend that can be understood qualitatively as follows.  For increasing noise amplitude, characteristic deviations from the mechanical equilibrium $x_1 = x_2 = 0$ also increase.  Trajectories induced by sufficiently strong noise preferentially sample the nonlinear part of the restoring force which is more confining than the linear part, and thus the stochastic area growth rate is inhibited relative to a purely linear restoring force.  The nonlinearity-enhanced confinement is also evident by comparing probability density histograms of linear and nonlinear cases driven by identical noise amplitudes, cf. Fig. \ref{fig6}. 

We now turn to analyzing the dependence of stochastic area upon noise amplitudes using the streamfunction.  In particular, we want to quantitatively understand the aforementioned confinement effect, whereby hard spring nonlinearity reduces the increase of heat transfer as the noise temperatures increase.  We work in the dimensionless formulation with $\mu_1 = \mu_2 = 1$ with dimensionless energy landscape given by (\ref{eq:potential_energy_nonlin}) and (\ref{eq:potential_energy_nonlin1}).  In order to implement a scaling argument we focus here on the case of weak coupling, $0 < \epsilon << 1$, and $u(x)$ homogeneous of degree $n > 0$, so that
\begin{equation}
u(ax) = a^n u(x),
\end{equation}
for all $x$ and $a > 0$.   We take the noise amplitudes to be
\begin{equation}
\sigma_1 = \sigma, \:  \sigma_2 = \sigma r,
\label{eq:rdef}
\end{equation}
where $\sigma$ is any positive constant and $r$ denotes the ratio of noise amplitudes $\sigma_2/\sigma_1$ \cite{Dannenberg_PRL_2014}.   For $\epsilon = 0$, the stationary probability density is proportional to the product of the effective Boltzmann factors of the two decoupled degrees of freedom,
\begin{equation}
\rho \sim \rho_0 \propto \exp\left(-\frac{u(x_1)}{\sigma_1^2}\right) \exp\left(-\frac{u(x_2)}{\sigma_2^2}\right).
\label{eq:Prob_density_leading}
\end{equation}
The leading $\epsilon \rightarrow 0$ approximation to the streamfunction $\psi(\mathbf{x})$ then satisfies   
\begin{equation}
-  \partial_1 \left(\frac{1}{\rho_0}\partial_1 \psi\right) - r^2\partial_2 \left(\frac{1}{\rho_0}\partial_2 \psi\right)  = \epsilon (1 - r^2).
\label{eq:Stream_BVP_leading}
\end{equation}
This equation is obtained by setting $\mu_1 = \mu_2 = 1$ in (\ref{eq:Stream_BVP1}), substituting (\ref{eq:rdef}) for $\sigma_1, \sigma_2$, replacing $\rho$ by $\rho_0$, and setting $\partial_{12} U = -\epsilon$ as follows from (\ref{eq:potential_energy_nonlin}).  Notice the cancellation of the factor $\sigma^2$ from both sides.  

In analogy with certain boundary layer problems in two-dimensional fluid systems, we note that the solutions of (\ref{eq:Stream_BVP_leading}) possess a scaling symmetry \cite{Neu_Singular_2015}.  Thus, letting $R(\mathbf{x})$ denote the probability density (\ref{eq:Prob_density_leading}) for $\sigma = 1$, we can express the probability density for any $\sigma > 0$ as 
\begin{equation}
\rho_0 (\mathbf{x}) = \frac{1}{\sigma^{4/n}} R\left(\bm{\xi} := \frac{\mathbf{x}}{\sigma^{2/n}}\right).
\end{equation}

Now, let $\Psi(\bm{\xi})$ denote the streamfunction for $\sigma = 1$.  We seek the stream function for any $\sigma > 0$ as
\begin{equation}
\psi(\mathbf{x}) = b \Psi(\bm{\xi}),
\label{eq:Psi_scaling_Ansatz}
\end{equation}
where the constant $b$ is to be determined. Substituting this form into (\ref{eq:Stream_BVP_leading}), we have
\begin{equation}
-b \left[ \frac {\partial}{\partial\xi_1} \left( \frac{1}{R} \frac {\partial\Psi}{\partial\xi_1} \right) + \sigma^2\frac {\partial}{\partial\xi_2} \left( \frac{1}{R} \frac {\partial\Psi}{\partial\xi_2}\right) \right] = \epsilon (1 - r^2),
\label{eq:Stream_BVP_scaled}
\end{equation}
which has the exact form of the original equation (\ref{eq:Stream_BVP_leading}) for $\psi$ when $\sigma = 1$, provided that we choose $b = 1$.  Hence, we conclude
\begin{equation}
\psi(\mathbf{x}) = \Psi \left(\frac{\mathbf{x}}{\sigma^{2/n}}\right).
\end{equation}
Using this result, we can now write the rate of change of stochastic area (\ref{eq:Stream_stoch_area}) as
\begin{equation}
\frac{d}{dt}\langle A \rangle = \int_{\mathbb{R}^2} \Psi\left(\frac{\mathbf{x}}{\sigma^{2/n}}\right) \mathbf{dx} = \sigma^{4/n} \int_{\mathbb{R}^2} \Psi(\bm{\xi}) d\bm{\xi}.
\end{equation}
and, hence, $d\langle A \rangle / dt$ scales as
\begin{equation}
\frac{d}{dt}\langle A \rangle \propto \sigma^{4/n}.
\end{equation}
For linear stochastic dynamics associated with a quadratic potential energy function $(n = 2)$, we have $d\langle A \rangle / dt \propto \sigma^2$, a quadratic dependence on noise amplitude.  However, for the case of quartic confining potential with $n = 4$, we find $d\langle A \rangle / dt \propto \sigma$, so that area growth grows only  \textit{linearly} with noise amplitude.  Recall the numerical graph of $d\langle A \rangle / dt$ versus $\sigma_1$ with $\sigma_2/\sigma_1 = 0.50$ in Fig. \ref{fig4} is apparently linear for $\sigma_1 > 1$.
Since temperatures $T_1$ and $T_2$ are proportional to $\sigma_1^2 = \sigma^2$ and $\sigma_2^2 = \sigma^2 r^2$, we rewrite (61) as
\begin{equation}
\frac{d}{dt}\langle A \rangle \propto (\sigma^2)^{2/n}.
\end{equation}
For linear stochatic dynamics with $n = 2$, we see that $d\langle A \rangle / dt$ increases \textit{linearly} with increasing temperatures, consistent with
(45).  For stiff nonlinearity with $n > 2$, the increase of $d\langle A \rangle / dt$ with temperatures is \textit{sublinear}.

%
%
%
%
\section{Conclusions}
\label{sec:Discussion}
A general framework has been presented for understanding the use of stochastic line integrals (SLIs) to quantitatively characterize dynamical behavior in noise-driven non-equilibrium systems.  The concept of SLI has been shown to apply both at the level of the full high-dimensional phase space associated with Hamiltonian flow as well as in the reduced phase space typically described by a SDE.  The implementation has been shown to be robust and numerical implementation schemes have been suggested.  We have also shown how general results play out in detail for a paradigmatic model system - two coupled mass-spring elements at different temperatures. Along the way, SLIs are developed that correspond to physically meaningful quantities such as heat flow. We have shown that the stochastic line integral of \textit{any} non-exact differental form vanishes in a detailed balance system, so its nonvanishing is a clear imprimatur of irreversiblity.  In summary, stochastic line integrals are robust indicators of irreversibility which can be calculated directly from experimental time series.  Collateral deductions of heat transfer (if appropriate) readily follow.  For many experimental systems (e.g., biophysical systems) this route may be more accessible than carrying out calorimetry.  

We have also shown how the average growth rate of stochastic line integrals on two-dimensional state spaces can be evaluated with the aid of stream functions.  While stream functions are limited to describing two dimensional systems, we note that for many higher dimensional systems one finds that essential irreversible behavior is captured by projecting onto two degrees for freedom.  In such cases, the stream function approach may be useful.  Regarding the evaluation of line integrals there are certainly straightforward approaches that do not require use of stream functions.  For example, one can solve the Fokker-Planck equation, extract the collateral probability current, and then determine the average growth of the desired stochastic line integral by explicit evaluation of 
(\ref{eq:Stoch_line_int_reduced}).  What then are some of the advantages to using a stream function approach? Among them, we have shown how the stream function formulation clearly connects to essential physical properties such as area growth rate and heat transport.  In the elliptic equation (\ref{eq:Stream_BVP2}) for $\psi$,   the source term on the RHS contains the product $(T_1 - T_2) \partial_{12}U$, so we see very clearly why temperature difference and actual coupling ($\partial_{12}U$ not identically zero) are essential for irreversibility.  Another benefit of the stream function formulation is that scaling arguments like the one at the end of Sec. \ref{sec:Nonlinearity} are vastly more accessible.

Finally, we point to one conspicuous direction for future work: the case of state-dependent noise.  There are physical contexts in which it appears unavoidable, such as circuits containing elements with nonlinear current-voltage relations \cite{Teitsworth_2019}.  Presumably, one still has recorded time series of state variables and its natural to ask:  do (suitably modified) stochastic line integrals of non-exact differential forms allow us to read irreversibility, and calculate heat transfer and entropy production in such situations?

\label{sec:Discussion}
%
%
%
%
\section{Acknowledgements}
We thank Konstantin Matveev for helpful comments concerning the manuscript.


\begin{thebibliography}{35}%
\makeatletter
\providecommand \@ifxundefined [1]{%
 \@ifx{#1\undefined}
}%
\providecommand \@ifnum [1]{%
 \ifnum #1\expandafter \@firstoftwo
 \else \expandafter \@secondoftwo
 \fi
}%
\providecommand \@ifx [1]{%
 \ifx #1\expandafter \@firstoftwo
 \else \expandafter \@secondoftwo
 \fi
}%
\providecommand \natexlab [1]{#1}%
\providecommand \enquote  [1]{``#1''}%
\providecommand \bibnamefont  [1]{#1}%
\providecommand \bibfnamefont [1]{#1}%
\providecommand \citenamefont [1]{#1}%
\providecommand \href@noop [0]{\@secondoftwo}%
\providecommand \href [0]{\begingroup \@sanitize@url \@href}%
\providecommand \@href[1]{\@@startlink{#1}\@@href}%
\providecommand \@@href[1]{\endgroup#1\@@endlink}%
\providecommand \@sanitize@url [0]{\catcode `\\12\catcode `\$12\catcode
  `\&12\catcode `\#12\catcode `\^12\catcode `\_12\catcode `\%12\relax}%
\providecommand \@@startlink[1]{}%
\providecommand \@@endlink[0]{}%
\providecommand \url  [0]{\begingroup\@sanitize@url \@url }%
\providecommand \@url [1]{\endgroup\@href {#1}{\urlprefix }}%
\providecommand \urlprefix  [0]{URL }%
\providecommand \Eprint [0]{\href }%
\providecommand \doibase [0]{http://dx.doi.org/}%
\providecommand \selectlanguage [0]{\@gobble}%
\providecommand \bibinfo  [0]{\@secondoftwo}%
\providecommand \bibfield  [0]{\@secondoftwo}%
\providecommand \translation [1]{[#1]}%
\providecommand \BibitemOpen [0]{}%
\providecommand \bibitemStop [0]{}%
\providecommand \bibitemNoStop [0]{.\EOS\space}%
\providecommand \EOS [0]{\spacefactor3000\relax}%
\providecommand \BibitemShut  [1]{\csname bibitem#1\endcsname}%
\let\auto@bib@innerbib\@empty
\bibitem [{\citenamefont {Battle}\ \emph {et~al.}(2016)\citenamefont {Battle},
  \citenamefont {Broedersz}, \citenamefont {Fakhri}, \citenamefont {Geyer},
  \citenamefont {Howard}, \citenamefont {Schmidt},\ and\ \citenamefont
  {MacKintosh}}]{Battle_Science_2016}%
  \BibitemOpen
  \bibfield  {author} {\bibinfo {author} {\bibfnamefont {Christopher}\
  \bibnamefont {Battle}}, \bibinfo {author} {\bibfnamefont {Chase~P.}\
  \bibnamefont {Broedersz}}, \bibinfo {author} {\bibfnamefont {Nikta}\
  \bibnamefont {Fakhri}}, \bibinfo {author} {\bibfnamefont {Veikko~F.}\
  \bibnamefont {Geyer}}, \bibinfo {author} {\bibfnamefont {Jonathon}\
  \bibnamefont {Howard}}, \bibinfo {author} {\bibfnamefont {Christoph~F.}\
  \bibnamefont {Schmidt}}, \ and\ \bibinfo {author} {\bibfnamefont {Fred~C.}\
  \bibnamefont {MacKintosh}},\ }\bibfield  {title} {\enquote {\bibinfo {title}
  {{Broken detailed balance at mesoscopic scales in active biological
  systems}},}\ }\href {\doibase {10.1126/science.aac8167}} {\bibfield
  {journal} {\bibinfo  {journal} {Science}\ }\textbf {\bibinfo {volume}
  {352}},\ \bibinfo {pages} {604--607} (\bibinfo {year} {2016})}\BibitemShut
  {NoStop}%
\bibitem [{\citenamefont {Gnesotto}\ \emph {et~al.}(2018)\citenamefont
  {Gnesotto}, \citenamefont {Mura}, \citenamefont {Gladrow},\ and\
  \citenamefont {Broedersz}}]{Gnesotto_2018}%
  \BibitemOpen
  \bibfield  {author} {\bibinfo {author} {\bibfnamefont {F.}~\bibnamefont
  {Gnesotto}}, \bibinfo {author} {\bibfnamefont {F.}~\bibnamefont {Mura}},
  \bibinfo {author} {\bibfnamefont {J.}~\bibnamefont {Gladrow}}, \ and\
  \bibinfo {author} {\bibfnamefont {C.~P.}\ \bibnamefont {Broedersz}},\
  }\bibfield  {title} {\enquote {\bibinfo {title} {Broken detailed balance and
  non-equilibrium dynamics in living systems: a review},}\ }\href@noop {}
  {\bibfield  {journal} {\bibinfo  {journal} {Rep. Prog. Phys.}\ }\textbf
  {\bibinfo {volume} {81}},\ \bibinfo {pages} {066601} (\bibinfo {year}
  {2018})}\BibitemShut {NoStop}%
\bibitem [{\citenamefont {Weiss}({2007})}]{Weiss_PRE_2007}%
  \BibitemOpen
  \bibfield  {author} {\bibinfo {author} {\bibfnamefont {Jeffrey~B.}\
  \bibnamefont {Weiss}},\ }\bibfield  {title} {\enquote {\bibinfo {title}
  {Fluctuation properties of steady-state langevin systems},}\ }\href@noop {}
  {\bibfield  {journal} {\bibinfo  {journal} {Phys. Rev. E}\ }\textbf {\bibinfo
  {volume} {{76}}},\ \bibinfo {pages} {061128} (\bibinfo {year}
  {{2007}})}\BibitemShut {NoStop}%
\bibitem [{\citenamefont {Weiss}\ \emph {et~al.}(2020)\citenamefont {Weiss},
  \citenamefont {Fox-Kemper}, \citenamefont {Mandal}, \citenamefont {Nelson},\
  and\ \citenamefont {Zia}}]{Weiss_JSP_2019}%
  \BibitemOpen
  \bibfield  {author} {\bibinfo {author} {\bibfnamefont {J.~B.}\ \bibnamefont
  {Weiss}}, \bibinfo {author} {\bibfnamefont {B.}~\bibnamefont {Fox-Kemper}},
  \bibinfo {author} {\bibfnamefont {D.}~\bibnamefont {Mandal}}, \bibinfo
  {author} {\bibfnamefont {A.~D.}\ \bibnamefont {Nelson}}, \ and\ \bibinfo
  {author} {\bibfnamefont {R.~K.~P.}\ \bibnamefont {Zia}},\ }\bibfield  {title}
  {\enquote {\bibinfo {title} {{Nonequilibrium Oscillations, Probability
  Angular Momentum, and the Climate System}},}\ }\href@noop {} {\bibfield
  {journal} {\bibinfo  {journal} {Journal of Statistical Physics}\ }\textbf
  {\bibinfo {volume} {{179}}},\ \bibinfo {pages} {1010} (\bibinfo {year}
  {2020})}\BibitemShut {NoStop}%
\bibitem [{\citenamefont {Gieseler}\ \emph {et~al.}(2014)\citenamefont
  {Gieseler}, \citenamefont {Quidant}, \citenamefont {Dellago},\ and\
  \citenamefont {Novotny}}]{Gieseler_2014}%
  \BibitemOpen
  \bibfield  {author} {\bibinfo {author} {\bibfnamefont {Jan}\ \bibnamefont
  {Gieseler}}, \bibinfo {author} {\bibfnamefont {Romain}\ \bibnamefont
  {Quidant}}, \bibinfo {author} {\bibfnamefont {Christoph}\ \bibnamefont
  {Dellago}}, \ and\ \bibinfo {author} {\bibfnamefont {Lukas}\ \bibnamefont
  {Novotny}},\ }\bibfield  {title} {\enquote {\bibinfo {title} {Dynamic
  relaxation of a levitated nanoparticle from a non-equilibrium steady
  state},}\ }\href@noop {} {\bibfield  {journal} {\bibinfo  {journal} {Nature
  Nanotechnology}\ }\textbf {\bibinfo {volume} {9}},\ \bibinfo {pages} {358}
  (\bibinfo {year} {2014})}\BibitemShut {NoStop}%
\bibitem [{\citenamefont {Millen}\ \emph {et~al.}(2014)\citenamefont {Millen},
  \citenamefont {Deesuwan}, \citenamefont {Barker},\ and\ \citenamefont
  {Anders}}]{Millen_2014}%
  \BibitemOpen
  \bibfield  {author} {\bibinfo {author} {\bibfnamefont {J.}~\bibnamefont
  {Millen}}, \bibinfo {author} {\bibfnamefont {T.}~\bibnamefont {Deesuwan}},
  \bibinfo {author} {\bibfnamefont {P.}~\bibnamefont {Barker}}, \ and\ \bibinfo
  {author} {\bibfnamefont {J.}~\bibnamefont {Anders}},\ }\bibfield  {title}
  {\enquote {\bibinfo {title} {{Nanoscale temperature measurements using
  non-equilibrium Brownian dynamics of a levitated nanosphere}},}\ }\href@noop
  {} {\bibfield  {journal} {\bibinfo  {journal} {Nature Nanotechnology}\
  }\textbf {\bibinfo {volume} {9}},\ \bibinfo {pages} {425} (\bibinfo {year}
  {2014})}\BibitemShut {NoStop}%
\bibitem [{\citenamefont {Gonzalez-Ballestero}\ \emph
  {et~al.}(2021)\citenamefont {Gonzalez-Ballestero}, \citenamefont
  {Aspelmeyer}, \citenamefont {Novotny}, \citenamefont {Quidant},\ and\
  \citenamefont {Romero-Isart}}]{GonzalezBallestero_Science_2021}%
  \BibitemOpen
  \bibfield  {author} {\bibinfo {author} {\bibfnamefont {C.}~\bibnamefont
  {Gonzalez-Ballestero}}, \bibinfo {author} {\bibfnamefont {M.}~\bibnamefont
  {Aspelmeyer}}, \bibinfo {author} {\bibfnamefont {L.}~\bibnamefont {Novotny}},
  \bibinfo {author} {\bibfnamefont {R.}~\bibnamefont {Quidant}}, \ and\
  \bibinfo {author} {\bibfnamefont {O.}~\bibnamefont {Romero-Isart}},\
  }\bibfield  {title} {\enquote {\bibinfo {title} {{Levitodynamics: Levitation
  and control of microscopic objects in vacuum}},}\ }\href {\doibase
  10.1126/science.abg3027} {\bibfield  {journal} {\bibinfo  {journal}
  {Science}\ }\textbf {\bibinfo {volume} {374}},\ \bibinfo {pages} {eabg3027}
  (\bibinfo {year} {2021})}\BibitemShut {NoStop}%
\bibitem [{\citenamefont {Bomze}\ \emph {et~al.}(2012)\citenamefont {Bomze},
  \citenamefont {Hey}, \citenamefont {Grahn},\ and\ \citenamefont
  {Teitsworth}}]{Bomze_PRL_2012}%
  \BibitemOpen
  \bibfield  {author} {\bibinfo {author} {\bibfnamefont {Yu.}\ \bibnamefont
  {Bomze}}, \bibinfo {author} {\bibfnamefont {R.}~\bibnamefont {Hey}}, \bibinfo
  {author} {\bibfnamefont {H.~T.}\ \bibnamefont {Grahn}}, \ and\ \bibinfo
  {author} {\bibfnamefont {S.~W.}\ \bibnamefont {Teitsworth}},\ }\bibfield
  {title} {\enquote {\bibinfo {title} {Noise-induced current switching in
  semiconductor superlattices: Observation of nonexponential kinetics in a
  high-dimensional system},}\ }\href {\doibase 10.1103/PhysRevLett.109.026801}
  {\bibfield  {journal} {\bibinfo  {journal} {Phys. Rev. Lett.}\ }\textbf
  {\bibinfo {volume} {109}},\ \bibinfo {pages} {026801} (\bibinfo {year}
  {2012})}\BibitemShut {NoStop}%
\bibitem [{\citenamefont {Gonzalez}\ \emph {et~al.}(2019)\citenamefont
  {Gonzalez}, \citenamefont {Neu},\ and\ \citenamefont
  {Teitsworth}}]{Gonzalez_PRE_2019}%
  \BibitemOpen
  \bibfield  {author} {\bibinfo {author} {\bibfnamefont {Juan~Pablo}\
  \bibnamefont {Gonzalez}}, \bibinfo {author} {\bibfnamefont {John~C.}\
  \bibnamefont {Neu}}, \ and\ \bibinfo {author} {\bibfnamefont {Stephen~W.}\
  \bibnamefont {Teitsworth}},\ }\bibfield  {title} {\enquote {\bibinfo {title}
  {Experimental metrics for detection of detailed balance violation},}\ }\href
  {\doibase 10.1103/PhysRevE.99.022143} {\bibfield  {journal} {\bibinfo
  {journal} {Phys. Rev. E}\ }\textbf {\bibinfo {volume} {99}},\ \bibinfo
  {pages} {022143} (\bibinfo {year} {2019})}\BibitemShut {NoStop}%
\bibitem [{\citenamefont {Teitsworth}\ \emph {et~al.}(2019)\citenamefont
  {Teitsworth}, \citenamefont {Olson},\ and\ \citenamefont
  {Bomze}}]{Teitsworth_2019}%
  \BibitemOpen
  \bibfield  {author} {\bibinfo {author} {\bibfnamefont {Stephen~W.}\
  \bibnamefont {Teitsworth}}, \bibinfo {author} {\bibfnamefont {Matthew~E.}\
  \bibnamefont {Olson}}, \ and\ \bibinfo {author} {\bibfnamefont {Yuriy}\
  \bibnamefont {Bomze}},\ }\bibfield  {title} {\enquote {\bibinfo {title}
  {Scaling properties of noise-induced switching in a bistable tunnel diode
  circuit},}\ }\href {\doibase {10.1140/epjb/e2019-90711-0}} {\bibfield
  {journal} {\bibinfo  {journal} {Eur. Phys. J. B}\ }\textbf {\bibinfo {volume}
  {92}},\ \bibinfo {pages} {74} (\bibinfo {year} {2019})}\BibitemShut {NoStop}%
\bibitem [{\citenamefont {Ciliberto}\ \emph
  {et~al.}(2013{\natexlab{a}})\citenamefont {Ciliberto}, \citenamefont
  {Imparato}, \citenamefont {Naert},\ and\ \citenamefont
  {Tanase}}]{Ciliberto_PRL_2013}%
  \BibitemOpen
  \bibfield  {author} {\bibinfo {author} {\bibfnamefont {S.}~\bibnamefont
  {Ciliberto}}, \bibinfo {author} {\bibfnamefont {A.}~\bibnamefont {Imparato}},
  \bibinfo {author} {\bibfnamefont {A.}~\bibnamefont {Naert}}, \ and\ \bibinfo
  {author} {\bibfnamefont {M.}~\bibnamefont {Tanase}},\ }\bibfield  {title}
  {\enquote {\bibinfo {title} {Heat flux and entropy produced by thermal
  fluctuations},}\ }\href {\doibase 10.1103/PhysRevLett.110.180601} {\bibfield
  {journal} {\bibinfo  {journal} {Phys. Rev. Lett.}\ }\textbf {\bibinfo
  {volume} {110}},\ \bibinfo {pages} {180601} (\bibinfo {year}
  {2013}{\natexlab{a}})}\BibitemShut {NoStop}%
\bibitem [{\citenamefont {Ciliberto}\ \emph
  {et~al.}(2013{\natexlab{b}})\citenamefont {Ciliberto}, \citenamefont
  {Imparato}, \citenamefont {Naert},\ and\ \citenamefont
  {Tanase}}]{Ciliberto_JSM_2013b}%
  \BibitemOpen
  \bibfield  {author} {\bibinfo {author} {\bibfnamefont {Sergio}\ \bibnamefont
  {Ciliberto}}, \bibinfo {author} {\bibfnamefont {Alberto}\ \bibnamefont
  {Imparato}}, \bibinfo {author} {\bibfnamefont {Antoine}\ \bibnamefont
  {Naert}}, \ and\ \bibinfo {author} {\bibfnamefont {Marius}\ \bibnamefont
  {Tanase}},\ }\bibfield  {title} {\enquote {\bibinfo {title} {Statistical
  properties of the energy exchanged between two heat baths coupled by thermal
  fluctuations},}\ }\href@noop {} {\bibfield  {journal} {\bibinfo  {journal}
  {Journal of Statistical Mechanics: Theory and Experiment}\ }\textbf {\bibinfo
  {volume} {2013}},\ \bibinfo {pages} {P12014} (\bibinfo {year}
  {2013}{\natexlab{b}})}\BibitemShut {NoStop}%
\bibitem [{\citenamefont {Fogedby}\ and\ \citenamefont
  {Imparator}(2012)}]{Fogedby_2012}%
  \BibitemOpen
  \bibfield  {author} {\bibinfo {author} {\bibfnamefont {Hans~C.}\ \bibnamefont
  {Fogedby}}\ and\ \bibinfo {author} {\bibfnamefont {Alberto}\ \bibnamefont
  {Imparator}},\ }\bibfield  {title} {\enquote {\bibinfo {title} {Heat flow in
  chains driven by thermal noise},}\ }\href@noop {} {\bibfield  {journal}
  {\bibinfo  {journal} {Journal of Statistical Mechanics: Theory and
  Experiment}\ }\textbf {\bibinfo {volume} {2012}},\ \bibinfo {pages} {P04005}
  (\bibinfo {year} {2012})}\BibitemShut {NoStop}%
\bibitem [{\citenamefont {Li}\ \emph {et~al.}(2019)\citenamefont {Li},
  \citenamefont {Horowitz}, \citenamefont {Gingrich},\ and\ \citenamefont
  {Fakhri}}]{Gingrich_2019}%
  \BibitemOpen
  \bibfield  {author} {\bibinfo {author} {\bibfnamefont {Junang}\ \bibnamefont
  {Li}}, \bibinfo {author} {\bibfnamefont {Jordan~M}\ \bibnamefont {Horowitz}},
  \bibinfo {author} {\bibfnamefont {Todd~R}\ \bibnamefont {Gingrich}}, \ and\
  \bibinfo {author} {\bibfnamefont {Nikta}\ \bibnamefont {Fakhri}},\ }\bibfield
   {title} {\enquote {\bibinfo {title} {Quantifying dissipation using
  fluctuating currents},}\ }\href@noop {} {\bibfield  {journal} {\bibinfo
  {journal} {Nat. Comm.}\ }\textbf {\bibinfo {volume} {10}},\ \bibinfo {pages}
  {1666} (\bibinfo {year} {2019})}\BibitemShut {NoStop}%
\bibitem [{\citenamefont {Zia}\ and\ \citenamefont
  {Schmittmann}(2007)}]{Zia_Schmittmann_2007}%
  \BibitemOpen
  \bibfield  {author} {\bibinfo {author} {\bibfnamefont {RKP}\ \bibnamefont
  {Zia}}\ and\ \bibinfo {author} {\bibfnamefont {B}~\bibnamefont
  {Schmittmann}},\ }\bibfield  {title} {\enquote {\bibinfo {title} {Probability
  currents as principal characteristics in the statistical mechanics of
  non-equilibrium steady states},}\ }\href@noop {} {\bibfield  {journal}
  {\bibinfo  {journal} {Journal of Statistical Mechanics: Theory and
  Experiment}\ }\textbf {\bibinfo {volume} {2007}},\ \bibinfo {pages} {P07012}
  (\bibinfo {year} {2007})}\BibitemShut {NoStop}%
\bibitem [{\citenamefont {Mellor}\ \emph {et~al.}(2016)\citenamefont {Mellor},
  \citenamefont {Mobilia},\ and\ \citenamefont {Zia}}]{Mellor_EPL_2016}%
  \BibitemOpen
  \bibfield  {author} {\bibinfo {author} {\bibfnamefont {A.}~\bibnamefont
  {Mellor}}, \bibinfo {author} {\bibfnamefont {M.}~\bibnamefont {Mobilia}}, \
  and\ \bibinfo {author} {\bibfnamefont {R.~K.~P.}\ \bibnamefont {Zia}},\
  }\bibfield  {title} {\enquote {\bibinfo {title} {{Characterization of the
  nonequilibrium steady state of a heterogeneous nonlinear q-voter model with
  zealotry}},}\ }\href@noop {} {\bibfield  {journal} {\bibinfo  {journal}
  {Europhys. Lett.}\ }\textbf {\bibinfo {volume} {113}},\ \bibinfo {pages}
  {48001} (\bibinfo {year} {2016})}\BibitemShut {NoStop}%
\bibitem [{\citenamefont {Gradziuk}\ \emph {et~al.}(2019)\citenamefont
  {Gradziuk}, \citenamefont {Mura},\ and\ \citenamefont
  {Broedersz}}]{Gradziuk_2019}%
  \BibitemOpen
  \bibfield  {author} {\bibinfo {author} {\bibfnamefont {Grzegorz}\
  \bibnamefont {Gradziuk}}, \bibinfo {author} {\bibfnamefont {Federica}\
  \bibnamefont {Mura}}, \ and\ \bibinfo {author} {\bibfnamefont {Chase~P}\
  \bibnamefont {Broedersz}},\ }\bibfield  {title} {\enquote {\bibinfo {title}
  {Scaling behavior of nonequilibrium measures in internally driven elastic
  assemblies},}\ }\href@noop {} {\bibfield  {journal} {\bibinfo  {journal}
  {Physical Review E}\ }\textbf {\bibinfo {volume} {99}},\ \bibinfo {pages}
  {052406} (\bibinfo {year} {2019})}\BibitemShut {NoStop}%
\bibitem [{\citenamefont {Gnesotto}\ \emph {et~al.}(2020)\citenamefont
  {Gnesotto}, \citenamefont {Gradziuk}, \citenamefont {Ronceray},\ and\
  \citenamefont {Broedersz}}]{Gnesotto_2020}%
  \BibitemOpen
  \bibfield  {author} {\bibinfo {author} {\bibfnamefont {Federico~S.}\
  \bibnamefont {Gnesotto}}, \bibinfo {author} {\bibfnamefont {Grzegorz}\
  \bibnamefont {Gradziuk}}, \bibinfo {author} {\bibfnamefont {Pierre}\
  \bibnamefont {Ronceray}}, \ and\ \bibinfo {author} {\bibfnamefont {Chase~P.}\
  \bibnamefont {Broedersz}},\ }\bibfield  {title} {\enquote {\bibinfo {title}
  {{Learning the non-equilibrium dynamics of Brownian movies}},}\ }\href@noop
  {} {\bibfield  {journal} {\bibinfo  {journal} {Nat. Comm.}\ }\textbf
  {\bibinfo {volume} {11}},\ \bibinfo {pages} {5378} (\bibinfo {year}
  {2020})}\BibitemShut {NoStop}%
\bibitem [{\citenamefont {Frishman}\ and\ \citenamefont
  {Ronceray}(2020)}]{Frishman_2020}%
  \BibitemOpen
  \bibfield  {author} {\bibinfo {author} {\bibfnamefont {Anna}\ \bibnamefont
  {Frishman}}\ and\ \bibinfo {author} {\bibfnamefont {Pierre}\ \bibnamefont
  {Ronceray}},\ }\bibfield  {title} {\enquote {\bibinfo {title} {{Learning
  force fields from stochastic trajectories}},}\ }\href {\doibase
  10.1103/PhysRevX.10.021009} {\bibfield  {journal} {\bibinfo  {journal} {Phys.
  Rev. X}\ }\textbf {\bibinfo {volume} {10}},\ \bibinfo {pages} {021009}
  (\bibinfo {year} {2020})}\BibitemShut {NoStop}%
\bibitem [{\citenamefont {Ghanta}\ \emph {et~al.}(2017)\citenamefont {Ghanta},
  \citenamefont {Neu},\ and\ \citenamefont {Teitsworth}}]{Ghanta2017}%
  \BibitemOpen
  \bibfield  {author} {\bibinfo {author} {\bibfnamefont {Akhil}\ \bibnamefont
  {Ghanta}}, \bibinfo {author} {\bibfnamefont {John~C}\ \bibnamefont {Neu}}, \
  and\ \bibinfo {author} {\bibfnamefont {Stephen}\ \bibnamefont {Teitsworth}},\
  }\bibfield  {title} {\enquote {\bibinfo {title} {{Fluctuation loops in
  noise-driven linear dynamical systems}},}\ }\href@noop {} {\bibfield
  {journal} {\bibinfo  {journal} {Physical Review E}\ }\textbf {\bibinfo
  {volume} {95}},\ \bibinfo {pages} {032128} (\bibinfo {year}
  {2017})}\BibitemShut {NoStop}%
\bibitem [{\citenamefont {Gardiner}(2009)}]{Gardiner_2009}%
  \BibitemOpen
  \bibfield  {author} {\bibinfo {author} {\bibfnamefont {C.}~\bibnamefont
  {Gardiner}},\ }\href {http://books.google.com/books?id=otg3PQAACAAJ} {\emph
  {\bibinfo {title} {Stochastic Methods: A Handbook for the Natural and Social
  Sciences}}},\ Springer Series in Synergetics\ (\bibinfo  {publisher}
  {Springer},\ \bibinfo {year} {2009})\BibitemShut {NoStop}%
\bibitem [{Note1()}]{Note1}%
  \BibitemOpen
  \bibinfo {note} {This also provides helpful insight when we apply metrics
  such as the stochastic line integral directly to real experimental data
  collected from high-dimensional systems. The process of experimentally
  measuring only a few macroscopic variables is qualitatively related to the
  theoretical process of projecting from a high-dimensional microscopic model
  (e.g., with number of degrees of freedom of order of Avogadgro's number) to
  SDE models described by (\ref {eq:SDE_general}).}\BibitemShut {Stop}%
\bibitem [{\citenamefont {Onsager}(1931)}]{Onsager_1931}%
  \BibitemOpen
  \bibfield  {author} {\bibinfo {author} {\bibfnamefont {Lars}\ \bibnamefont
  {Onsager}},\ }\bibfield  {title} {\enquote {\bibinfo {title} {Reciprocal
  relations in irreversible processes. i.}}\ }\href@noop {} {\bibfield
  {journal} {\bibinfo  {journal} {Phys. Rev.}\ }\textbf {\bibinfo {volume}
  {37}},\ \bibinfo {pages} {405} (\bibinfo {year} {1931})}\BibitemShut
  {NoStop}%
\bibitem [{\citenamefont {Van~Kampen}(2007)}]{vanKampen_2007}%
  \BibitemOpen
  \bibfield  {author} {\bibinfo {author} {\bibfnamefont {N.~G.}\ \bibnamefont
  {Van~Kampen}},\ }\href@noop {} {\emph {\bibinfo {title} {Stochastic processes
  in physics and chemistry}}},\ \bibinfo {edition} {3rd}\ ed.\ (\bibinfo
  {publisher} {Elsevier},\ \bibinfo {year} {2007})\BibitemShut {NoStop}%
\bibitem [{\citenamefont {Zwanzig}(2001)}]{Zwanzig_2001}%
  \BibitemOpen
  \bibfield  {author} {\bibinfo {author} {\bibfnamefont {Robert}\ \bibnamefont
  {Zwanzig}},\ }\href@noop {} {\emph {\bibinfo {title} {Nonequilibrium
  statistical mechanics}}}\ (\bibinfo  {publisher} {Oxford University Press},\
  \bibinfo {year} {2001})\BibitemShut {NoStop}%
\bibitem [{\citenamefont {Bennett}(1982)}]{Bennett_1982}%
  \BibitemOpen
  \bibfield  {author} {\bibinfo {author} {\bibfnamefont {Charles~H}\
  \bibnamefont {Bennett}},\ }\bibfield  {title} {\enquote {\bibinfo {title}
  {The thermodynamics of computation—a review},}\ }\href@noop {} {\bibfield
  {journal} {\bibinfo  {journal} {International Journal of Theoretical
  Physics}\ }\textbf {\bibinfo {volume} {21}},\ \bibinfo {pages} {905--940}
  (\bibinfo {year} {1982})}\BibitemShut {NoStop}%
\bibitem [{\citenamefont {Tolman}(1938)}]{Tolman1938}%
  \BibitemOpen
  \bibfield  {author} {\bibinfo {author} {\bibfnamefont {Richard~C.}\
  \bibnamefont {Tolman}},\ }\href@noop {} {\emph {\bibinfo {title} {The
  principles of statistical mechanics}}}\ (\bibinfo  {publisher} {Courier
  Corporation},\ \bibinfo {year} {1938})\BibitemShut {NoStop}%
\bibitem [{\citenamefont {Luchinsky}\ \emph {et~al.}(1998)\citenamefont
  {Luchinsky}, \citenamefont {McClintock},\ and\ \citenamefont
  {Dykman}}]{Luchinsky_RPP_1998}%
  \BibitemOpen
  \bibfield  {author} {\bibinfo {author} {\bibfnamefont {D~G}\ \bibnamefont
  {Luchinsky}}, \bibinfo {author} {\bibfnamefont {P~V~E}\ \bibnamefont
  {McClintock}}, \ and\ \bibinfo {author} {\bibfnamefont {M~I}\ \bibnamefont
  {Dykman}},\ }\bibfield  {title} {\enquote {\bibinfo {title} {Analogue studies
  of nonlinear systems},}\ }\href
  {http://stacks.iop.org/0034-4885/61/i=8/a=001} {\bibfield  {journal}
  {\bibinfo  {journal} {Rep. Prog. Phys.}\ }\textbf {\bibinfo {volume} {61}},\
  \bibinfo {pages} {905} (\bibinfo {year} {1998})}\BibitemShut {NoStop}%
\bibitem [{Note2()}]{Note2}%
  \BibitemOpen
  \bibinfo {note} {More generally, it less clear that one might infer this
  separation of time scales looking directly at the structure of the full $H$.
  Related to this point, there are certainly many Hamiltonian flows for which
  this form of separation of time scales is not expected \cite
  {Zwanzig_2001}.}\BibitemShut {Stop}%
\bibitem [{\citenamefont {Chiang}\ \emph
  {et~al.}(2017{\natexlab{a}})\citenamefont {Chiang}, \citenamefont {Lee},
  \citenamefont {Lai},\ and\ \citenamefont {Chen}}]{Chiang_2017a}%
  \BibitemOpen
  \bibfield  {author} {\bibinfo {author} {\bibfnamefont {K-H}\ \bibnamefont
  {Chiang}}, \bibinfo {author} {\bibfnamefont {C-L}\ \bibnamefont {Lee}},
  \bibinfo {author} {\bibfnamefont {P-Y}\ \bibnamefont {Lai}}, \ and\ \bibinfo
  {author} {\bibfnamefont {Y-F}\ \bibnamefont {Chen}},\ }\bibfield  {title}
  {\enquote {\bibinfo {title} {Entropy production and irreversibility of
  dissipative trajectories in electric circuits},}\ }\href@noop {} {\bibfield
  {journal} {\bibinfo  {journal} {Physical Review E}\ }\textbf {\bibinfo
  {volume} {95}},\ \bibinfo {pages} {012158} (\bibinfo {year}
  {2017}{\natexlab{a}})}\BibitemShut {NoStop}%
\bibitem [{\citenamefont {Chiang}\ \emph
  {et~al.}(2017{\natexlab{b}})\citenamefont {Chiang}, \citenamefont {Lee},
  \citenamefont {Lai},\ and\ \citenamefont {Chen}}]{Chiang_2017b}%
  \BibitemOpen
  \bibfield  {author} {\bibinfo {author} {\bibfnamefont {K-H}\ \bibnamefont
  {Chiang}}, \bibinfo {author} {\bibfnamefont {C-L}\ \bibnamefont {Lee}},
  \bibinfo {author} {\bibfnamefont {P-Y}\ \bibnamefont {Lai}}, \ and\ \bibinfo
  {author} {\bibfnamefont {Y-F}\ \bibnamefont {Chen}},\ }\bibfield  {title}
  {\enquote {\bibinfo {title} {{Electrical autonomous Brownian gyrator}},}\
  }\href@noop {} {\bibfield  {journal} {\bibinfo  {journal} {Physical Review
  E}\ }\textbf {\bibinfo {volume} {96}},\ \bibinfo {pages} {032123} (\bibinfo
  {year} {2017}{\natexlab{b}})}\BibitemShut {NoStop}%
\bibitem [{\citenamefont {Neu}(2009)}]{Neu_Training_2009}%
  \BibitemOpen
  \bibfield  {author} {\bibinfo {author} {\bibfnamefont {J.C.}\ \bibnamefont
  {Neu}},\ }\href {https://books.google.com/books?id=0IuDAwAAQBAJ} {\emph
  {\bibinfo {title} {Training Manual on Transport and Fluids}}},\ Graduate
  studies in mathematics\ (\bibinfo  {publisher} {American Mathematical
  Society},\ \bibinfo {year} {2009})\BibitemShut {NoStop}%
\bibitem [{\citenamefont {Neu}(2015)}]{Neu_Singular_2015}%
  \BibitemOpen
  \bibfield  {author} {\bibinfo {author} {\bibfnamefont {J.C.}\ \bibnamefont
  {Neu}},\ }\href {https://books.google.com/books?id=rv8dCwAAQBAJ} {\emph
  {\bibinfo {title} {Singular Perturbation in the Physical Sciences}}},\
  Graduate Studies in Mathematics\ (\bibinfo  {publisher} {American
  Mathematical Society},\ \bibinfo {year} {2015})\BibitemShut {NoStop}%
\bibitem [{\citenamefont {Evans}(2010)}]{Evans_PDE_2010}%
  \BibitemOpen
  \bibfield  {author} {\bibinfo {author} {\bibfnamefont {L.C.}\ \bibnamefont
  {Evans}},\ }\href {https://books.google.com/books?id=Xnu0o\_EJrCQC} {\emph
  {\bibinfo {title} {Partial Differential Equations}}},\ Graduate studies in
  mathematics\ (\bibinfo  {publisher} {American Mathematical Society},\
  \bibinfo {year} {2010})\BibitemShut {NoStop}%
\bibitem [{\citenamefont {Dannenberg}\ \emph {et~al.}({2014})\citenamefont
  {Dannenberg}, \citenamefont {Neu},\ and\ \citenamefont
  {Teitsworth}}]{Dannenberg_PRL_2014}%
  \BibitemOpen
  \bibfield  {author} {\bibinfo {author} {\bibfnamefont {Paul~H.}\ \bibnamefont
  {Dannenberg}}, \bibinfo {author} {\bibfnamefont {John~C.}\ \bibnamefont
  {Neu}}, \ and\ \bibinfo {author} {\bibfnamefont {Stephen~W.}\ \bibnamefont
  {Teitsworth}},\ }\bibfield  {title} {\enquote {\bibinfo {title} {Steering
  most probable escape paths by varying relative noise intensities},}\ }\href
  {\doibase {10.1103/PhysRevLett.113.020601}} {\bibfield  {journal} {\bibinfo
  {journal} {{Phys. Rev. Lett.}}\ }\textbf {\bibinfo {volume} {{113}}},\
  \bibinfo {pages} {020601} (\bibinfo {year} {{2014}})}\BibitemShut {NoStop}%
\end{thebibliography}
\end{document}